\documentclass[preprint,review,12pt]{elsarticle}

\usepackage{subcaption}
\usepackage{amssymb}
\usepackage{amsmath}
\usepackage{xcolor}
\usepackage{booktabs}
\usepackage{graphicx}
\usepackage{hyperref}
\usepackage{lineno}

\journal{Ocean Engineering}

\begin{document}

\begin{frontmatter}

\title{Sloshing suppression with a controlled elastic baffle via deep reinforcement learning and SPH simulation}

\author[inst1]{Mai Ye}

\author[inst2]{Yaru Ren\texorpdfstring{\corref{cor1}}{}}
\ead{renyaru818@163.com}

\author[inst3]{Silong Zhang}

\author[inst4]{Hao Ma}

\author[inst1]{Xiangyu Hu\texorpdfstring{\corref{cor1}}{}}
\ead{xiangyu.hu@tum.de}

\author[inst1]{Oskar J. Haidn}

\cortext[cor1]{Corresponding author}

\affiliation[inst1]{organization={TUM School of Engineering and Design, Technical University of Munich},
            postcode={85748}, 
            city={Garching},
            country={Germany}}

\affiliation[inst2]{organization={State Key Laboratory of Hydraulics and Mountain River Engineering, Sichuan University}, 
            postcode={610065}, 
            city={Chengdu},
            country={China}}

\affiliation[inst3]{organization={School of Energy Science and Engineering, Harbin Institute of Technology}, 
            postcode={150001}, 
            city={Harbin},
            country={China}}

\affiliation[inst4]{organization={School of Aerospace Engineering, Zhengzhou University of Aeronautics}, 
            postcode={450046}, 
            city={Zhengzhou},
            country={China}}

\begin{abstract}
This study employed smoothed particle hydrodynamics (SPH) as the numerical environment, integrated with deep reinforcement learning (DRL) real-time control algorithms to optimize the sloshing suppression in a tank with a centrally positioned vertical elastic baffle. Compared to rigid baffle movement and active strain control methods, the active-controlled movable elastic baffle, which remains undeformed at its base, achieved the best performance with an 81.63\% reduction in mean free surface amplitude. A cosine-based expert policy derived from DRL data is also extracted, resulting in a comparable 76.86\% reduction in a three-dimensional (3D) numerical simulation. Energy analyses showed that elastic baffle motion effectively decreased system energy by performing negative work on the fluid, reducing kinetic and potential energy. The DRL-based and expert policies also demonstrated robust performance across varying excitation frequencies and water depths. Specifically, rigid baffles proved more effective at frequencies below the system's first natural frequency, while elastic baffles exhibited superior performance at higher frequencies. Changes in water depth minimally affected the effectiveness of control policies, though they significantly influenced elastic baffle deformation behavior. Overall, the sloshing suppression efficiency consistently ranged between 70\% and 80\%, confirming DRL-informed control methods' versatility and effectiveness under diverse operating conditions.
\end{abstract}


\end{frontmatter}

\section{Introduction}
In partially filled containers, the free surface of the liquid is often disturbed by the motion of the container, leading to sloshing. This phenomenon is commonly observed in various engineering applications, including fluid-transporting tank trucks \cite{kolaei2015three}, marine liquefied natural gas (LNG) carriers \cite{lee2007effects}, spacecraft fuel tanks \cite{veldman2007numerical, kotsarinis2023modeling}, and aircraft wing-tank sloshing\cite{nerattini2024loads}. With the rapid advancement of computer technology, the past decades have witnessed significant progress in numerical and experimental studies on sloshing mechanisms \cite{ibrahim2005liquid}. \citet{faltinsen2001adaptive} developed a multimodal method for calculating nonlinear sloshing in two-dimensional (2D) rectangular tanks with finite water depth. Building on this work, they extended their research to resonant nonlinear sloshing in three-dimensional (3D) square-base basins, analyzing the influence of tank dimensions and excitation modes \cite{faltinsen2003resonant, faltinsen2017resonant}. Many researchers also studied resonant sloshing in square-based tanks using numerical simulations \cite{liu2008numerical, ikeda2012nonlinear}. More recently, \citet{bardazzi2024different} employed an enhanced smoothed particle hydrodynamics (SPH) method to analyze horizontally oscillating sloshing flows near the critical filling depth. Additionally, several authors conducted experimental investigations, focusing on laboratory measurements of water depth and hydrodynamic pressures during sloshing \cite{souto2011set, pistani2012experimental}. It is worth noting that when the external excitation frequency approaches the resonant frequency or exhibits a large amplitude, the liquid inside the container can undergo intense oscillations \cite{faltinsen2021coupling}. This results in significant impact forces on the container walls, potentially leading to structural fatigue, wear, and potential damage over time. While baffles are widely regarded as one of the most practical and effective methods for mitigating sloshing \cite{zheng2021review}, the interaction between sloshing and baffles constitutes a complex fluid-structure interaction (FSI) problem, which is highly nonlinear and challenging to model accurately.

Research on baffles can be broadly categorized into two main types: fixed and moving baffles. The studies further classify fixed baffles based on their placement and structural design. Placement configurations include vertical \cite{jung2012effect}, horizontal \cite{sanapala2018numerical}, annular \cite{gavrilyuk2006sloshing}, and multiple baffles \cite{xue2012numerical}, while structural designs encompass rectangular \cite{liu2009three}, T-shaped \cite{unal2019liquid}, porous \cite{cho2017sloshing}, and nonlinear baffles \cite{raja2023numerical}. In the past decade, numerous scholars have conducted research involving theoretical analysis, numerical simulations, and experimental studies. \citet{hasheminejad2014liquid} employed linear potential theory to solve the Laplace equation and analyzed the sloshing behavior of liquid in a circular tank equipped with baffles under the assumptions of an inviscid, incompressible fluid and irrotational flow. \citet{cho2017sloshing} employed the matched eigenfunction expansion method (MEEM) to explore the analytical solutions of the sloshing problem in a rectangular tank with horizontal porous baffles, aiming to identify the optimal porosity under various operating conditions. \citet{akyildiz2012numerical} investigated the impact of vertical baffles on liquid sloshing in a 2D rectangular tank using the finite difference method (FDM) and the volume of fluid (VOF) technique to solve Navier–Stokes and continuity equations. \citet{ma2021numerical} conducted a simulation analysis on the suppression effects of single and double vertical baffles on intense sloshing under resonant conditions with finite water depth. Their study employed the lattice Boltzmann method (LBM) combined with the VOF technique and large eddy simulation (LES). \citet{xue2017experimental} experimentally investigated the effectiveness of four different types of baffles in suppressing sloshing over a wide range of external excitation frequencies. \citet{chu2018slosh} found that as the number of vertical baffles increases, the tank's water depth and hydrodynamic forces decrease through experiments.

In recent years, research on moving baffles for suppressing sloshing has gradually become a new focus of interest. It can be broadly categorized into two main types: studies focusing on moving rigid baffles \cite{pan2018design} and those on elastic baffles \cite{idelsohn2008interaction}. The first approach involves using floating vertical baffles for moving baffles \cite{bellezi2022numerical, lu2024experimental}. This method operates on a principle similar to oscillating float wave energy generators, converting wave energy into kinetic energy, increasing wave energy dissipation, and effectively reducing sloshing \cite{sun2021structural}. The second approach connects the baffle to a spring system, where the spring absorbs the large impact forces from the liquid, thereby improving the system's sloshing response \cite{kim2018experimental, gligor2024sloshing}. \citet{gligor2024microgravity} employed passive spring-mass baffles to suppress impulsive sloshing in microgravity environments. Numerical results indicated that movable baffles significantly improved sloshing mitigation compared to fixed baffles, reducing the damping time by up to 48\%. As an alternative to spring systems, elastic baffles absorb kinetic energy from the liquid through baffle deformation \cite{zhang2020investigations, xue2023two}. \citet{ren2023numerical, ren2023experimental} were the first to conduct a detailed experimental investigation of tank sloshing with a vertical elastic baffle, complemented by a 2D numerical study using the SPH method. Their findings demonstrated that elastic baffles exhibit superior sloshing suppression capabilities compared to rigid baffles within specific water depths and external excitation frequency ranges. Additionally, elastic baffles effectively balance the impact forces on the tank walls and the baffles.

It is worth noting that the aforementioned studies primarily focus on passive sloshing suppression using baffles. Research on actively controlled baffles is relatively scarce. Among the few studies, \citet{xie2021sloshing, xie2022active} applied deep reinforcement learning (DRL) to actively control the vertical displacement of horizontal baffles placed near the tank walls in real time to suppress sloshing. Their results demonstrated that sloshing could be reduced by up to 81.48\%, highlighting that DRL-based active flow control (AFC) methods are arguably a promising solution for sloshing suppression. 
 
DRL is a machine learning method that combines reinforcement learning (RL) \cite{sutton2018reinforcement} with deep learning (DL). The core concept of RL is rooted in the Markov decision process (MDP) and dynamic programming (DP), where an agent interacts with the environment through trial and error to learn an optimal decision-making policy for low-dimensional problems. With the rapid development of deep neural networks (DNNs) \cite{sze2017efficient}, integrating DNNs with classical RL algorithms, such as Q-learning \cite{watkins1992q}, has enabled DRL to identify optimal policies in higher-dimensional and continuous spaces. This advancement has significantly expanded the application of DRL in various engineering fields \cite{silver2017mastering, aradi2020survey}. In the field of fluid mechanics, DRL is primarily used for structural optimization \cite{viquerat2021direct, keramati2022deep} and AFC \cite{hachem2021deep, vignon2023recent}. \citet{rabault2019artificial} were the first to use DRL to control the drag characteristics of flow around a cylinder. They simulated 2D cylinder flow at a moderate Reynolds number ($Re = 100$), and by controlling the flow rates at two nozzles positioned above and below the cylinder, they achieved an 8\% reduction in the drag coefficient while maintaining a nearly constant lift coefficient. Following their study, extensive research on drag reduction around bluff bodies has been conducted \cite{paris2021robust, wang2024dynamic}. In addition, using DRL to explore optimal swimming strategies for underwater locomotion has become another research hotspot \cite{verma2018efficient, wang2024learn}.

Currently, there is no research on actively controlled elastic baffles for sloshing suppression. This study addresses this gap by building upon previous work \cite{ren2023numerical} and employing DRL to control elastic vertical baffles. The control strategies will primarily involve moving the baffles and applying active strains. Considering this is a typical FSI problem, the SPH method is particularly advantageous for solving issues involving large structural movements or deformations \cite{zhang2021multi}. Therefore, the numerical simulations will be implemented using the open-source SPH multi-physics library SPHinXsys \cite{zhang2021sphinxsys}. The accuracy of the numerical model will be validated through previous experiments \cite{ren2023experimental} and integrated with the DRL platform Tianshou \cite{weng2022tianshou} to enable real-time simulation and control. The rest of the paper is organized as follows. Section~\ref{NM} primarily focuses on establishing the physical and numerical model. Section~\ref{RLAF} the coupling of the RL algorithms. Section~\ref{Results} presents the DRL training results and investigates the sloshing suppression mechanisms, including the impact of the 3D numerical model, control strategies, different external excitation frequencies and water depths. In Section~\ref{Conclusions}, the summary and conclusions are provided.

\section{Numerical modeling}\label{NM}
\subsection{Physical model}
\begin{figure}
  \centerline{\includegraphics[scale=0.55]{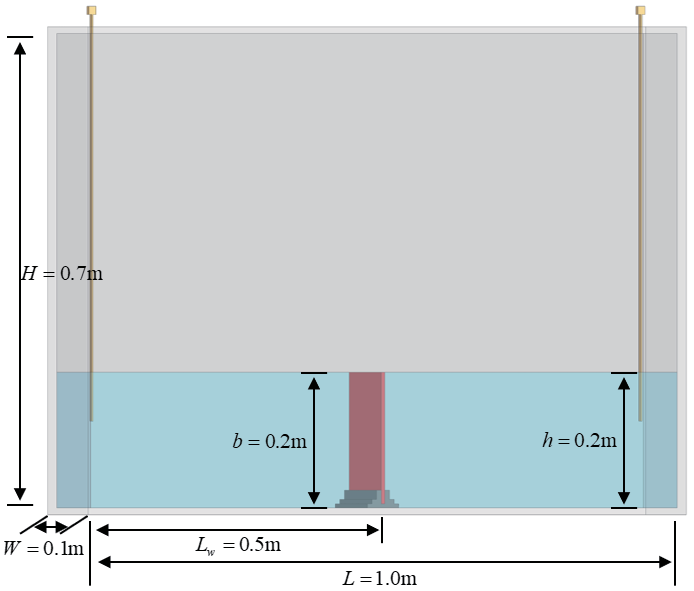}}
  \caption{The sketch of the sloshing experiment.}
\label{fig:1}
\end{figure}

We first constructed the physical model based on the structure used in previous experiments \cite{ren2023experimental}, as shown in Fig.~\ref{fig:1}. The tank dimensions are \(1.0 \, \mathrm{m} \times 0.7 \, \mathrm{m} \times 0.1 \, \mathrm{m}\), with an elastic baffle fixed at the center of the tank using a slot. Both the baffle and the slot have a width of \(0.095 \, \mathrm{m}\), ensuring no contact with the tank walls. The height of the baffle from the top to the tank bottom is \(b = 0.2 \, \mathrm{m}\). The baffle material is polyurethane with the following properties: Young's modulus \(E = 30 \, \mathrm{MPa}\), density \(\rho_s = 1250 \, \mathrm{kg/m}^3\), and Poisson's ratio \(\nu^s = 0.47\). The water depth is defined as \(h\). Two free surface probes are positioned on the left and right sides, each located \(0.032 \, \mathrm{m}\) from the tank walls. 

This study primarily considers horizontal periodic sinusoidal external excitations. The acceleration \(\mathbf{a}^{e}\) acting on the fluid and baffle is expressed as \cite{liu2009three}:
\begin{equation}
a^e_x = -A\omega_e^2\sin(\omega_e t),
\label{eq:1}
\end{equation}
where \(A\) represents the excitation amplitude, and \(\omega_e = 2 \pi f_e\) denotes the angular frequency of the excitation. 

\begin{figure}
  \centerline{\includegraphics[scale=0.7]{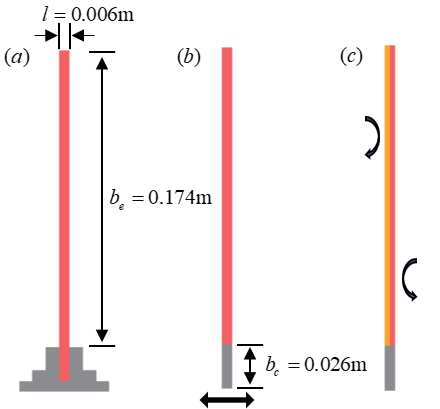}}
  \caption{The configuration of the connection between the baffle and the slot $(a)$. Controlled horizontal translation of the constrained region at the base of the baffle to induce overall baffle movement $(b)$. Application of active strain on both sides of the baffle while keeping the constrained region at the base stationary $(c)$.}
\label{fig:2}
\end{figure}

In this study, both elastic and rigid baffles are employed, with their geometric configurations kept identical. Moreover, We assume that the deformation of the elastic baffle constrained by the slot is negligible, and the shape of the slot has little effect on the interaction between the baffle and the sloshing. Thus, the constrained region can be simplified as a rigid component for RL training, as shown in Fig.~\ref{fig:2}. Two approaches are chosen for controlling the deformation of the elastic baffle. The first method applies horizontal movement to the constrained region, similar to the approach used by \citet{xie2021sloshing}, which controls the vertical motion of a rigid horizontal baffle. The baffle undergoes only passive deformation resulting from interacting with the fluid. The second method keeps the constrained region stationary and divides the baffle into two parts, each with a thickness of \(0.003 \, \mathrm{m}\). Active strain is applied to both parts to induce the baffle's active bending. This approach is inspired by the work of \citet{curatolo2016modeling}, who successfully modeled the active distortions of red muscles in fish to simulate swimming. The specific equations for the control can be found in Section \ref{RLAF}.

\subsection{Governing equations}
\subsubsection{Fluid model}
Tank sloshing can be considered an incompressible viscous flow \cite{liu2009three}. Within the SPH framework, the Lagrangian formulations of the continuity equation and the momentum equation are:
\begin{equation}
\left\{
\begin{aligned}
\frac{d\rho}{dt} & = -\rho\nabla \cdot \mathbf{v},\\
\frac{d\mathbf{v}}{dt} & = -\frac{1}{\rho}\nabla p + \nu\nabla^2\mathbf{v} + \mathbf{a}^{sf} + \mathbf{a}^{e} + \mathbf{g},
\end{aligned}
\right.
\label{eq:2}
\end{equation}
where $\rho$, $\mathbf{v}$, $p$, and $\nu$ are the fluid density, velocity, pressure, and kinematic viscosity, respectively. \(\mathbf{g}\) is the gravitational acceleration, $\mathbf{a}^{sf}$ the acceleration from structure. An artificial equation of state is used to close Eq.~(\ref{eq:2}):
\begin{equation}
\label{eq:3}
p = c^2(\rho - \rho^{0}).
\end{equation}
Here, $\rho^{0}$ is the initial density and $c = 10u_{max}$ is the artificial sound speed, where $u_{max} = 2 \sqrt{|\mathbf{g}|h}$ is the maximum predictable particle velocity in the flow \cite{zhang2021sphinxsys}, $h$ the water depth. 

The discrete of Eq.~(\ref{eq:2}) can be written as:
\begin{equation}
\left\{
\begin{aligned}
\frac{d\rho_{i}}{dt} & = 
    2\rho_{i}\sum_{j}(\mathbf{v}_{i} - \mathbf{v}^{*})\nabla W_{ij}{V_{j}},\\
\frac{d\mathbf{v}_{i}}{dt} & = 
    -\frac{2}{\rho_{i}}\sum_{j}P^*\nabla W_{ij}V_{j} + \frac{2}{\rho_{i}}\sum_{j}\mu\frac{\mathbf{v}_{ij}}{r_{ij}}\frac{\partial W_{ij}}{\partial r_{ij}}V_{j} + \mathbf{a}^{sf}_{i} + \mathbf{a}^{e}_{i} + \mathbf{g}_{i},
\end{aligned}
\right.
\label{eq:4}
\end{equation}
where $\nabla W_{ij} =\mathbf{e}_{ij} ( \partial W({r}_{ij}, h) / \partial r_{ij})$, $\mathbf{r}_{ij} = \mathbf{r}_{i} - \mathbf{r}_{j}$, $\mathbf{e}_{ij} = \mathbf{r}_{ij} / r_{ij}$, $\mathbf{v}_{ij} = \mathbf{v}_{i} - \mathbf{v}_{j}$ the particle relative velocity. $V_{j}$ and $\rho_{i}$ are the volume and density of corresponding particles, $\mu$ the dynamic viscosity. $\mathbf{v}^*$ and $P^*$ are the solutions of the one-dimensional Riemann problem which can be described as:
\begin{equation}
\left\{
\begin{aligned}
(\rho_{L}, U_{L}, P_{L}, c_{L}) & = 
    (\rho_{i}, -\mathbf{v}_{i} \cdot \mathbf{e}_{ij}, p_{i}, c_{i}),\\
(\rho_{R}, U_{R}, P_{R}, c_{R}) & = 
    (\rho_{j}, -\mathbf{v}_{j} \cdot \mathbf{e}_{ij}, p_{j}, c_{j}).
\end{aligned}
\right.
\label{eq:5}
\end{equation}
With a linearized Riemann solver \cite{zhang2017weakly} and the weighted kernel gradient correction \cite{ren2023efficient}, $\mathbf{v}^*$ and $P^*$ can be calculated as:
\begin{equation}
\left\{
\begin{aligned}
\mathbf{v}^* & = 
    \frac{\mathbf{v}_{i} + \mathbf{v}_{j}}{2} + (\frac{\rho_{L}c_{L}U_{L} + \rho_{R}c_{R}U_{R} + P_{L} - P_{R}}{\rho_{L}c_{L} + \rho_{R}c_{R}} - \frac{U_{L} + U_{R}}{2})\mathbf{e}_{ij},\\
P^{*} & = 
    \frac{\rho_{L}c_{L}P_{R}\Bar{\mathbb{B}}_j + \rho_{R}c_{R}P_{L}\Bar{\mathbb{B}}_i + \rho_{L}c_{L}\rho_{R}c_{R}\beta(U_{L} - U_{R})}{\rho_{L}c_{L} + \rho_{R}c_{R}}.
\end{aligned}
\right.
\label{eq:6}
\end{equation}
Here, $\Bar{\mathbb{B}}_i = \omega_1 \mathbb{B}_i + \omega_2 \mathbb{I}$, $\omega_1 = |\mathbb{A}_i| / (|\mathbb{A}_i| + \epsilon)$, $\omega_2 = \epsilon / (|\mathbb{A}_i| + \epsilon)$, $\mathbb{B}_i = (\mathbb{A}_i)^{-1}$, $\mathbb{I}$ the identity matrix, $\epsilon = \max(\alpha - |\mathbb{A}_i|, 0)$, $\mathbb{A}_i = -\sum_{j}\mathbf{r}_{ij} \otimes \nabla W_{ij} V_{j}$ and $\alpha = 0.5$.

The dual time-stepping approach, which sets advection criterion $\Delta t_{ad}$ for the updates of particle configurations and the smaller acoustic criterion $\Delta t_{ac}$ for the pressure and density relaxation, is used here to enhance computational efficiency \citep{zhang2021sphinxsys}:
\begin{equation}
\left\{
\begin{aligned}
\Delta t_{ad} & = 
    CFL_{ad}\min(\frac{h_f}{u_{max}}, \frac{h_f^2}{\nu}),\\
\Delta t_{ac} & = 
    CFL_{ac}(\frac{h_f}{c + u_{max}}),
\end{aligned}
\right.
\label{eq:7}
\end{equation}
where $CFL_{ad} = 0.25$, $CFL_{ac} = 0.6$, $h_f$ the smoothing length.

\subsubsection{Solid model}
The Lagrangian formulations of the mass and momentum conservation equations for the linear and isotropic elastic structure are:
\begin{equation}
\left\{
\begin{aligned}
\rho^s & = 
    \rho^s_0\frac{1}{\det(\mathbb{F})},\\
\frac{d\mathbf{v}^s}{dt} & = 
    \frac{1}{\rho^s} \nabla_0 \cdot \mathbb{P}^{T} + \mathbf{a}^{fs} + \mathbf{a}^e + \mathbf{g}.
\end{aligned}
\right.
\label{eq:8}
\end{equation}
where $\rho^s_0$ the initial structure density, $\mathbb{F} = \nabla_0(\mathbf{r}^s - \mathbf{r}^s_0) + \mathbb{I}$ is the deformation gradient tensor. $\mathbb{P} = \mathbb{F}\mathbb{S}$ represents the first Piola–Kirchhoff stress tensor, $\mathbb{S}$ the second Piola–Kirchhoff stress tensor with Saint Venant–Kirchhoff model:
\begin{equation}
\label{eq:9}
\mathbb{S} = \frac{E \nu^s}{(1 + \nu^s)(1-2\nu^s)}\text{tr}(\mathbb{E}) + 2\frac{E}{2(1 + \nu^s)}\mathbb{E}.
\end{equation}
Here, $\mathbb{E} = (\mathbb{F}^{T}\mathbb{F} - \mathbb{I}) / 2$ is the Green–Lagrange strain tensor.

The discrete of (\ref{eq:8}) can be written as:
\begin{equation}
\left\{
\begin{aligned}
  \rho^s_i & = 
    \rho^s_0\frac{1}{\det((- \sum_{j} \mathbf{r}_{ij}^s \nabla_0 W_{ij}^s V_j^s)\mathbb{B}_{i}^0 + \mathbb{I})},\\
  \frac{d\mathbf{v}^s_i}{dt} & = 
    \frac{1}{\rho^s_i} \sum_j (\mathbb{P}_i \mathbb{B}_{i}^0 + \mathbb{P}_j \mathbb{B}_{j}^0) \nabla_0 W_{ij}^s V_j^s + \mathbf{a}^{fs}_i + \mathbf{a}^e_i + \mathbf{g}_i,
\end{aligned}
\right.
\label{eq:10}
\end{equation}
where $\mathbb{B}_{i}^0 = (- \sum_{j} (\mathbf{r}_{ij}^s)_0 \nabla_0 W_{ij}^s V_j^s)^{-1}$ is the kernel gradient correction matrix \cite{zhang2021multi}.

For the application of active strain on both sides of the baffle, the deformation gradient tensor is modified as $\mathbb{F}_t = \mathbb{F} \mathbb{F}_a$, where $\mathbb{F}_a$ represents a time-varying tensor. The corresponding Green strain tensor for the active component is changed to $\mathbb{E}_a = (\mathbb{F}_a^{T}\mathbb{F}_a - \mathbb{I}) / 2$. Consequently, the first Piola–Kirchhoff stress tensor, $\mathbb{P}_t$, is updated to $\mathbb{P}_t = \mathbb{P} \mathbb{F}_a^*$, where $\mathbb{F}_a^* = \det(\mathbb{F}_a)(\mathbb{F}_a^{-1})^T$. This formulation accounts for the influence of the active strain field on the overall deformation and stress distribution in the baffle.

Additionally, the time step for structure is
\begin{equation}
\label{eq:11}
\Delta t^{s} = 0.6\min(\frac{h^s}{c^s + u_{max}}, \sqrt{\frac{h^s}{(du/dt)_{max}}}),
\end{equation}
where $c^s = \sqrt{(\lambda^s +2 \mu^s/3)/\rho^s}$ represents the structure sound speed.

\subsubsection {Fluid-structure interaction}
In SPHinXsys, the solid boundary is treated as a dynamic wall boundary and the FSI problem can still be modeled as a one-sided Riemann problem:
\begin{equation}
\left\{
\begin{aligned}
  (\rho_{L}, U_{L}, P_{L}) & = 
    (\rho_{i}, -\mathbf{v}_{i} \cdot \mathbf{n}_{a}, p_{i}),\\
  (\rho_{R}, U_{R}, P_{R}) & = 
    (\rho_{a}, -(2\mathbf{v}_{i}-\Bar{\mathbf{v}}_a) \cdot \mathbf{n}_{a}, p_{a}),
\end{aligned}
\right.
\label{eq:12}
\end{equation}
where $a$ the solid particle, $\mathbf{v}_{ia} = \mathbf{v}_i - \Bar{\mathbf{v}}_a$,  $\Bar{\mathbf{v}}_a$ is the averaged velocity of the solid particle within $\Delta t_{ac}$. With non-slip boundary assumption, $p_{a} = p_{i} + \rho_i \max(0, (\mathbf{g} - d\Bar{\mathbf{v}}_a / dt)\cdot \mathbf{n}_{a})(\mathbf{r}_{ia}\cdot\mathbf{n}_{a})$, and the total force from structure $\mathbf{a}^{sf}$ can be written as:
\begin{equation}
\label{eq:13}
\mathbf{a}^{sf} = -\frac{2}{\rho_i}\sum_a P^* \nabla W_{ia} V_a + \frac{2}{\rho_i}\sum_a \mu \frac{\mathbf{v}_{ia}}{r_{ia}}\frac{\partial W_{ia}}{\partial r_{ia}}V_{a}.
\end{equation}

\subsection{Validations}
\begin{figure}[htbp]
    \centering
    \begin{subfigure}{\textwidth}
        \centering
        \includegraphics[width=\textwidth]{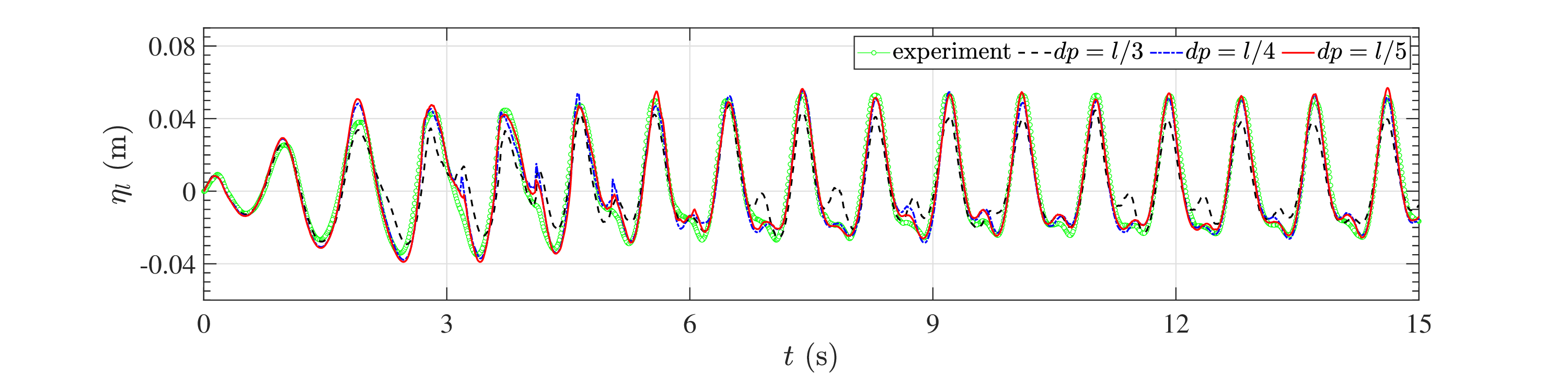}
        \caption{elastic baffle, \(h = 0.15 \, \mathrm{m}\), \(A = 0.01 \, \mathrm{m}\), \(f_e = 1.1 \, \mathrm{Hz}\)}
    \end{subfigure}
    
    \begin{subfigure}{\textwidth}
        \centering
        \includegraphics[width=\textwidth]{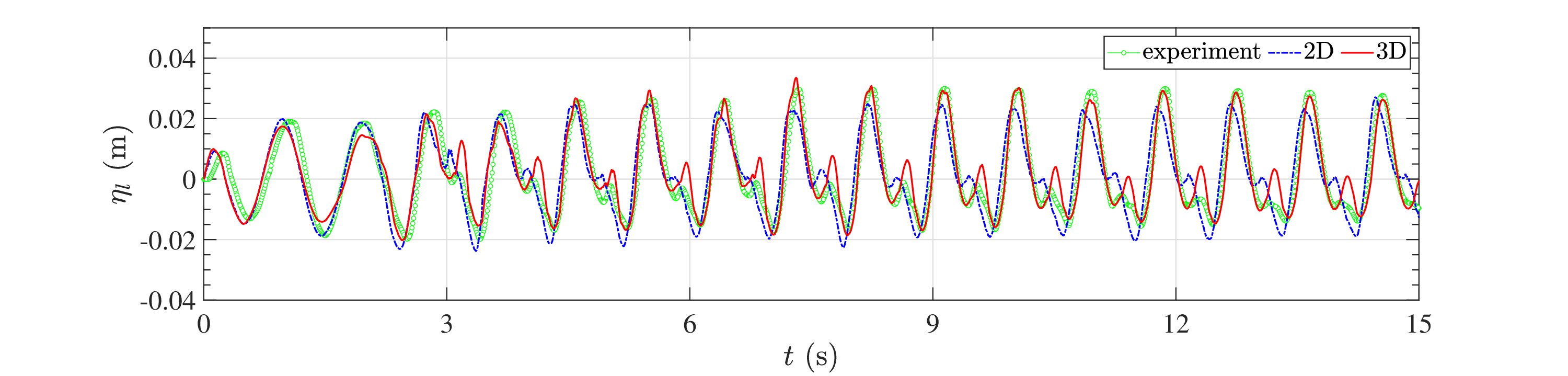}
        \caption{elastic baffle, \(h = 0.2 \, \mathrm{m}\), \(A = 0.01 \, \mathrm{m}\), \(f_e = 1.1 \, \mathrm{Hz}\)}
    \end{subfigure}

    \begin{subfigure}{\textwidth}
        \centering
        \includegraphics[width=\textwidth]{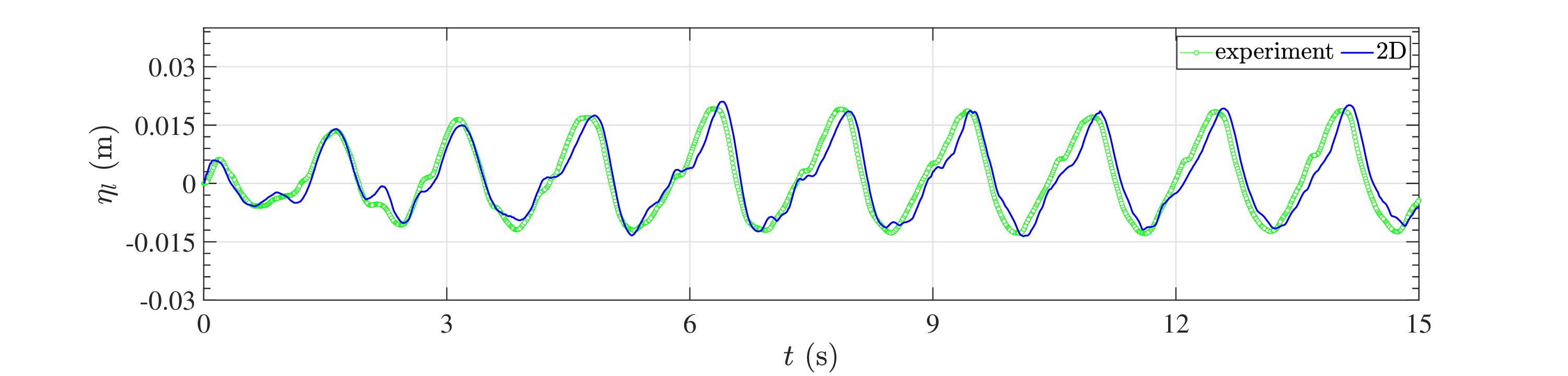}
        \caption{rigid baffle, \(h = 0.2 \, \mathrm{m}\), \(A = 0.01 \, \mathrm{m}\), \(f_e = 1.1 \, \mathrm{Hz}\)}
    \end{subfigure}
    
    \begin{subfigure}{\textwidth}
        \centering
        \includegraphics[width=\textwidth]{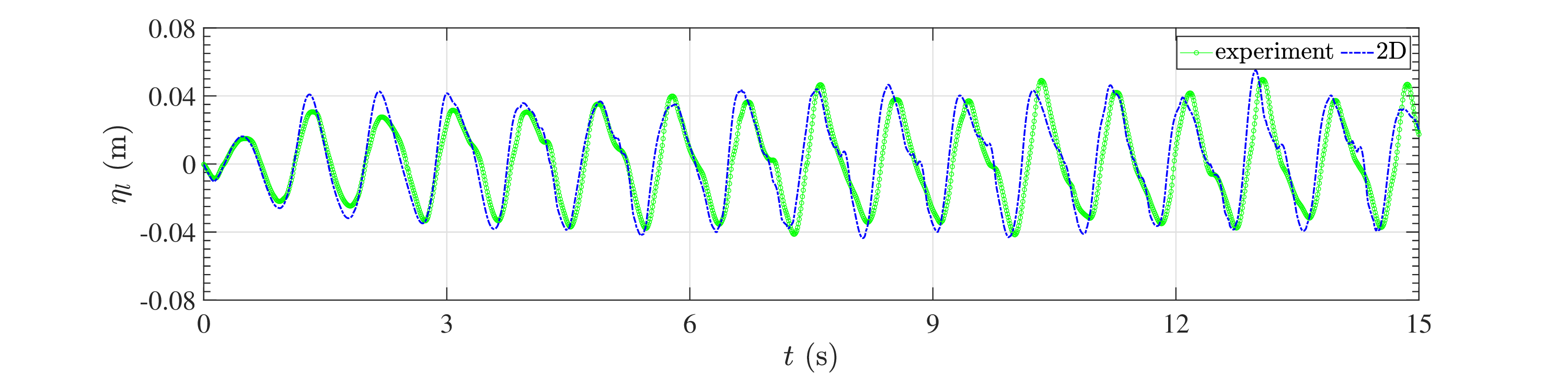}
        \caption{rigid baffle, \(h = 0.25 \, \mathrm{m}\), \(A = 0.01 \, \mathrm{m}\), \(f_e = 0.64 \, \mathrm{Hz}\)}
    \end{subfigure}

    \caption{Convergence analysis and experimental comparison of the free surface height at left wall \cite{ren2023experimental}.}
    \label{fig:3}
\end{figure}

This section presents a comprehensive analysis of model convergence and experimental validation for 2D and 3D simulations. For the 2D model, a baffle thickness of \(l = 0.006 \, \mathrm{m}\) is considered, with four layers of particles distributed along the thickness direction, yielding a structural resolution of \(dp_s = l/4\). As shown in Fig.~\ref{fig:3} (a), the convergence analysis of fluid particles reveals that the predicted free surface height near the left wall remains highly consistent across simulations using resolutions of \(dp = l/4\) and \(dp = l/5\), both exhibiting strong agreement with experimental observations. As a result, a fluid particle resolution of \(dp = l/4\) is adopted for all subsequent 2D simulations in this study. Moreover, the numerical simulations can capture the free surface height variations under different water depths (0.15 m, 0.2 m, and 0.25 m) and external excitation frequencies (0.64 Hz and 1.1 Hz), indicating the robustness of this study's numerical method and applicability across a broad frequency range. 

It is also worth noting that 3D simulations provide a more accurate representation than 2D simulations with the same particle resolution, as shown in Fig.~\ref{fig:3} (b). The 3D simulation accurately captures the variations in free surface height. In contrast, while the 2D simulation maintains periodic consistency, it exhibits a certain deviation in free surface height. Due to computational constraints, the numerical simulations were conducted on an AMD 3990X processor (64 cores and 128 processing threads) with 128 GB RAM, where each training episode in the 2D simulation environment, corresponding to 10 seconds of physical time, required approximately 0.25 hours to complete. In contrast, the 3D simulations took around 7.0 hours per episode. Regardless of the specific DRL algorithm used, the agent generally needs to collect at least 300 episodes of data to update both the value and policy networks, directly employing a 3D numerical environment as the training setup can be prohibitively time-consuming and inefficient. Therefore, in this study, 2D numerical simulations were utilized during the training phase to ensure computational feasibility. However, Section~\ref{3deffects} also includes a comparative analysis of the performance of the 2D-trained policy when applied to the 3D numerical environment.

\section{Reinforcement learning algorithms and framework}\label{RLAF}
\begin{figure}
  \centerline{\includegraphics[scale=0.5]{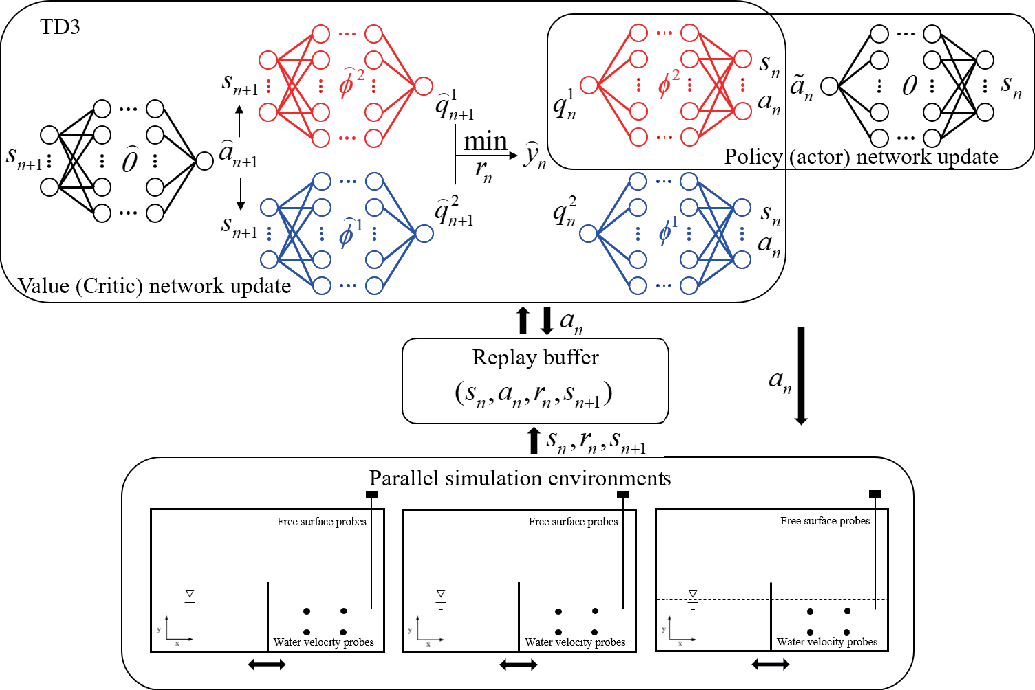}}
  \caption{The sketch of the DRL framework, which consists of three components. The SPH simulation environment, written in \texttt{C++}, is wrapped into a dynamic link library using Pybind11. This enables seamless data exchange and training between the parallel simulation environment and the DRL algorithm and replay buffer, which are implemented in \texttt{Python}.}
\label{fig:4}
\end{figure}

We treat the baffle as the agent in the decision-making process and formulate this as a Markov decision process (MDP). From Fig.~\ref{fig:4}, we can see that the agent interacts with the numerical simulation environment using an action policy generated by the RL algorithm, collecting data that includes the current and subsequent states \(s_n\), \(s_{n+1}\), actions \(a_n\), and rewards \(r_n\). The state is composed of three components: (1) the height of the free surface at various positions within the tank, (2) the velocity of the fluid at different locations inside the tank, and (3) the deformation and motion of the baffle, represented by the displacement along the \(y\)-direction at different positions relative to its initial configuration. A total of 42 measurement probes are selected, yielding 65 observations as state \(s_n\), and \(s_n\) was normalized prior to being fed into the neural network. The detailed convergence analysis results regarding the number of state sensors are in \ref{app1}. 

As mentioned before, the specific forms of the action can be categorized into two types. One of these involves the horizontal movement of the bottom constraint region where $a_n = \Delta u_x$ represents the change in the baffle's horizontal velocity \(u_x\) during an action duration of \(t_a = 0.05 \, \mathrm{s}\), with the restrictions \(|\Delta u_x| \leq 0.1 \, \mathrm{m/s}\) and \(|u_x| \leq 0.2 \, \mathrm{m/s}\). Another approach involves applying active strain $\mathbb{E}_a$ in the vertical direction to both sides of the baffle:
\begin{equation}
\label{eq:14}
  \mathbf{E}^{(1,1)}_a(y) = \left\{
    \begin{array}{ll}
      -\chi (1 - e^{-2t})\frac{b^2 - y^2}{b^2}, & x\le 0.5 L, \\
      -(0.06 - \chi) (1 - e^{-2t})\frac{b^2 - y^2}{b^2}, & x>0.5 L.
    \end{array} \right.
\end{equation}
The action is defined as \(a_n = \Delta \chi\), with constraints \(|\Delta \chi| \leq 0.01\) and \(|\chi| \leq 0.06\). The initial value of \(u_x\) and \(\chi\) are \(0.15 \ \mathrm{m/s}\) and 0.03, respectively. It is worth noting that the actions are linearly interpolated into the numerical simulation environment to maintain computational stability and prevent divergence. Each training episode consists of 200 action time steps, beginning at a physical time of \(t = 20\, \mathrm{s}\). This setup ensures that the baffle and the sloshing interaction have reached a stable dynamic regime before controlled actions are applied.

The reward \(r_n\) can be defined based on the free surface height at the left and right walls:
\begin{equation}
\label{eq:15}
r_n = \eta_b - \max(|\eta_l - h|, |\eta_r - h|).
\end{equation}
Here, \(\eta_b\) represents the maximum free surface height in the case without any control policy applied.

The data tuple \((s_n, a_n, r_n, s_{n+1})\) is used to update the DNNs within the actor-critic architecture of the RL algorithm, thereby optimizing the policy as shown in Fig.~\ref{fig:4}. In this framework, the actor usually serves as the policy network $\pi_\theta$, while the critic functions as the value network $q_{\phi}$. The policy network typically outputs a probability density function of actions, where a Gaussian distribution is constructed by predicting the mean and variance for sampling used in algorithms such as proximal policy optimization (PPO) and soft actor-critic (SAC) \cite{haarnoja2018soft} or direct deterministic action values with exploring noise, as seen in twin delayed deep deterministic policy gradient (TD3) \cite{fujimoto2018addressing}. The value network, on the other hand, can represent either an action-value function $q_{\pi_\theta}(s_{n}, a_{n}) = \mathbf{E}_{\pi_\theta}[\sum_{t=n}^{\infty} (\gamma_{t} r_{t} | s_{t}, a_{t})]$, which estimates the expected return with the discount factor $\gamma_{t} \in [0, 1]$ for a given state-action pair $(s_n, a_n)$, or a state-value function $V_{\pi_\theta}(s_{n}) = \mathbf{E}_{a_{n} \sim \pi_\theta}[\sum_{t=n}^{\infty} (\gamma_{t} r_{t} | s_{t})]$, which evaluates the expected return for a specific state $s_{n}$ regardless of the action taken.

The optimization and update of the policy network are usually achieved by constructing an objective function \( J(\theta) \) involving the parameters of both the policy and value networks. At this stage, the parameters of the value network are typically not updated. Instead, the data tuple \((s_n, a_n, r_n, s_{n+1})\) is used with gradient ascent to maximize the function, thereby updating \(\theta\). In contrast, the value network is updated by formulating a loss function $L(\phi_i)$ based on the temporal-difference (TD) method and applying gradient descent to minimize it with $\phi$. In this paper, we primarily use the TD3 algorithm, where the objective function and loss function can be expressed as:
\begin{equation}
\left\{
\begin{aligned}
  J(\theta) & = 
    \mathbf{E}_{s_{n} \sim \mathcal{B}} \left[ q_{\phi_1}(s_{n}, \pi_{\theta}(s_{n})) \right],\\
  L(\phi_i) & = 
    \mathbf{E}_{s_{n} \sim \mathcal{B}} \left[ \left( q_{\phi_i}(s_{n}, a_{n}) - y_{n} \right)^2 \right], i = 1, 2.
\end{aligned}
\right.
\label{eq:16}
\end{equation}
Here, \(\mathcal{B}\) denotes the replay buffer. For off-policy algorithms such as TD3, the replay buffer stores all previously collected data and samples it based on prioritized experience replay (PER) to update the networks. Additionally, the target value $y_{n}$ is defined as:
\begin{equation}
\label{eq:17}
y_n = r_n + \gamma \min_{i=1,2} q_{\widetilde{\phi}_i}(s_{n+1}, \pi_{\widetilde{\theta}}(s_{n+1}) + \epsilon),
\end{equation}
where $\epsilon$ is the truncated Gaussian noise. Clipped double Q-learning is employed to address the overestimation problem in Q-learning. \(\widetilde{\phi}_i\) and \(\widetilde{\theta}\) represent the parameters of the target networks. Detailed results of the algorithm comparison, including PPO, SAC, and TD3, are also provided in \ref{app1}.

\section{Results}\label{Results}
\subsection{Sloshing suppression through an active-controlled elastic baffle}\label{3deffects}
\begin{figure}[htbp]
    \centering
    \begin{subfigure}{\textwidth}
        \centering
        \includegraphics[width=\textwidth]{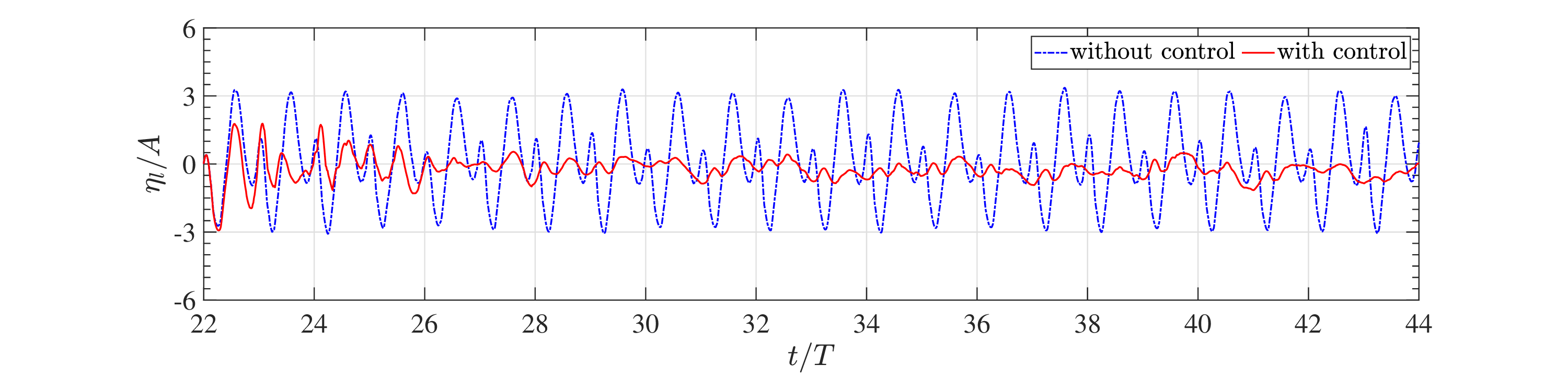}
        \caption{free surface height}
    \end{subfigure}
    
    \begin{subfigure}{\textwidth}
        \centering
        \includegraphics[width=\textwidth]{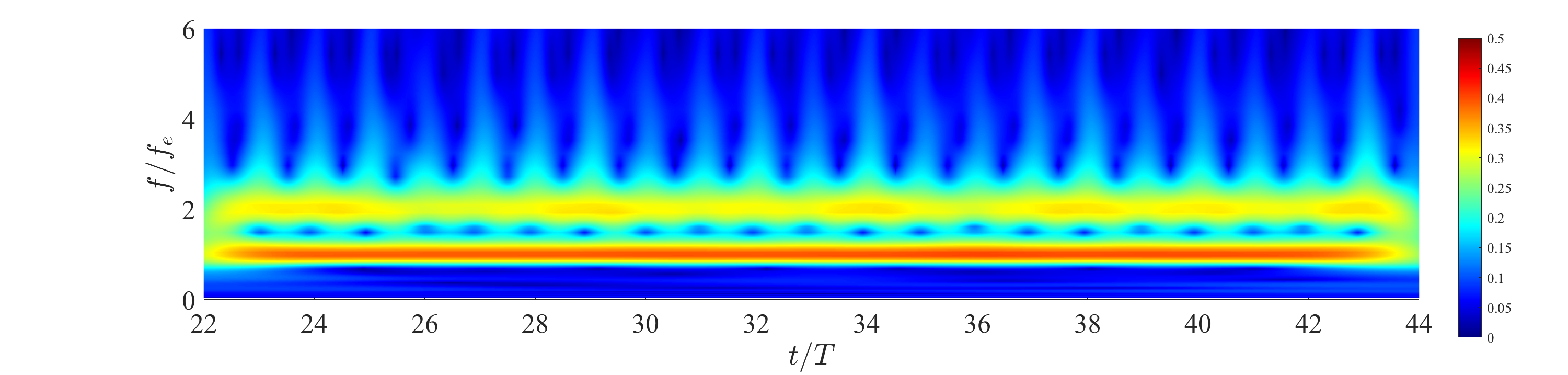}
        \caption{without control}
    \end{subfigure}
    
    \begin{subfigure}{\textwidth}
        \centering
        \includegraphics[width=\textwidth]{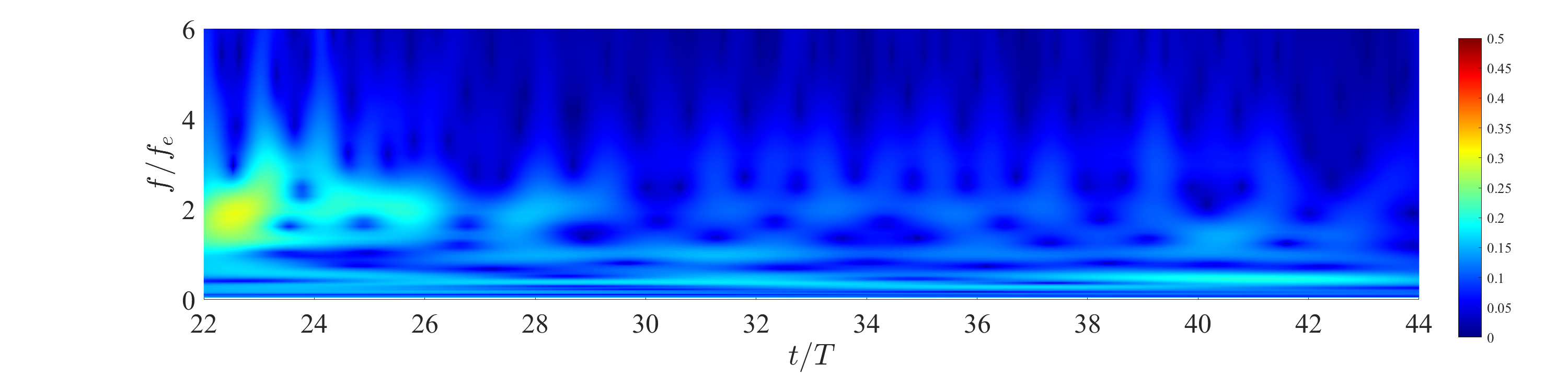}
        \caption{with control}
    \end{subfigure}

    \caption{Comparison of the free surface height and corresponding wavelet transforms at the left wall: (a) time history, (b) without control, and (c) with control.}
    \label{fig:5}
\end{figure}

The case with a water depth \( h = 0.2 \, \mathrm{m} \) and an external excitation frequency \( f_e = 1.1 \, \mathrm{Hz} \) is selected as the baseline scenario for training. The control policy is trained from the 22nd to the 33rd wave period (starting at \( t = 20 \, \mathrm{s} \)) after the system reaches a statistically steady state. Testing is conducted from the 22nd to the 44th period. As shown in Fig.~\ref{fig:5} (a), active control significantly suppresses sloshing at the left tank wall. Without control, the mean free surface amplitude $\eta_{mean}$ over 10 periods after stability is \( 0.0147 \, \mathrm{m} \), which decreases to \( 0.0027 \, \mathrm{m} \) with control, representing a 81.63\% reduction. 

Moreover, \citet{faltinsen2009sloshing} proposed that the natural angular frequency \( \omega_0 = 2\pi f_0 \) of a tank with an immersed bottom-mounted baffle can be estimated by
\begin{equation}
\label{eq:18}
\frac{\omega_0}{\omega} = 1 - \frac{2\pi^2\sin^2{\left(\pi n (x_0+0.5L)/L\right)}}{\sinh{\left(2\pi nh/L\right)}}, \quad n = 1, 2, \dots
\end{equation}
where \( \omega^2 = |\mathbf{g}|k\tanh(kh) \) represents the dispersion relation and \( k \) is the wave number. Here, \( x_0 \) denotes the position at which the baffle is constrained. Since the baffle moves laterally within the tank, the effective natural frequency of the system varies accordingly, directly influencing the sloshing dynamics. This behavior is reflected directly in the wavelet transform shown in Fig.~\ref{fig:5} (b) and (c). When the baffle remains stationary, the dominant frequency of the free surface height is \( 1.1 \, \mathrm{Hz} \), which aligns with both the external excitation and the theoretical first natural frequency \( f_0 = 1.1 \, \mathrm{Hz} \). A second harmonic at \( 2.2 \, \mathrm{Hz} \) is also observed, attributed to large-amplitude surface motion and nonlinear FSI characteristics. Once the baffle begins to move, the natural frequency shifts, and the system enters a transient regime lasting approximately four-wave periods. Thereafter, the free surface no longer exhibits a precise dominant frequency. A minor spectral peak emerges after the 38th wave period at around \( 0.45 \, \mathrm{Hz} \), suggesting a new quasi-equilibrium state under the modified dynamic conditions.

\begin{figure}
\centering
\includegraphics[width=\textwidth]{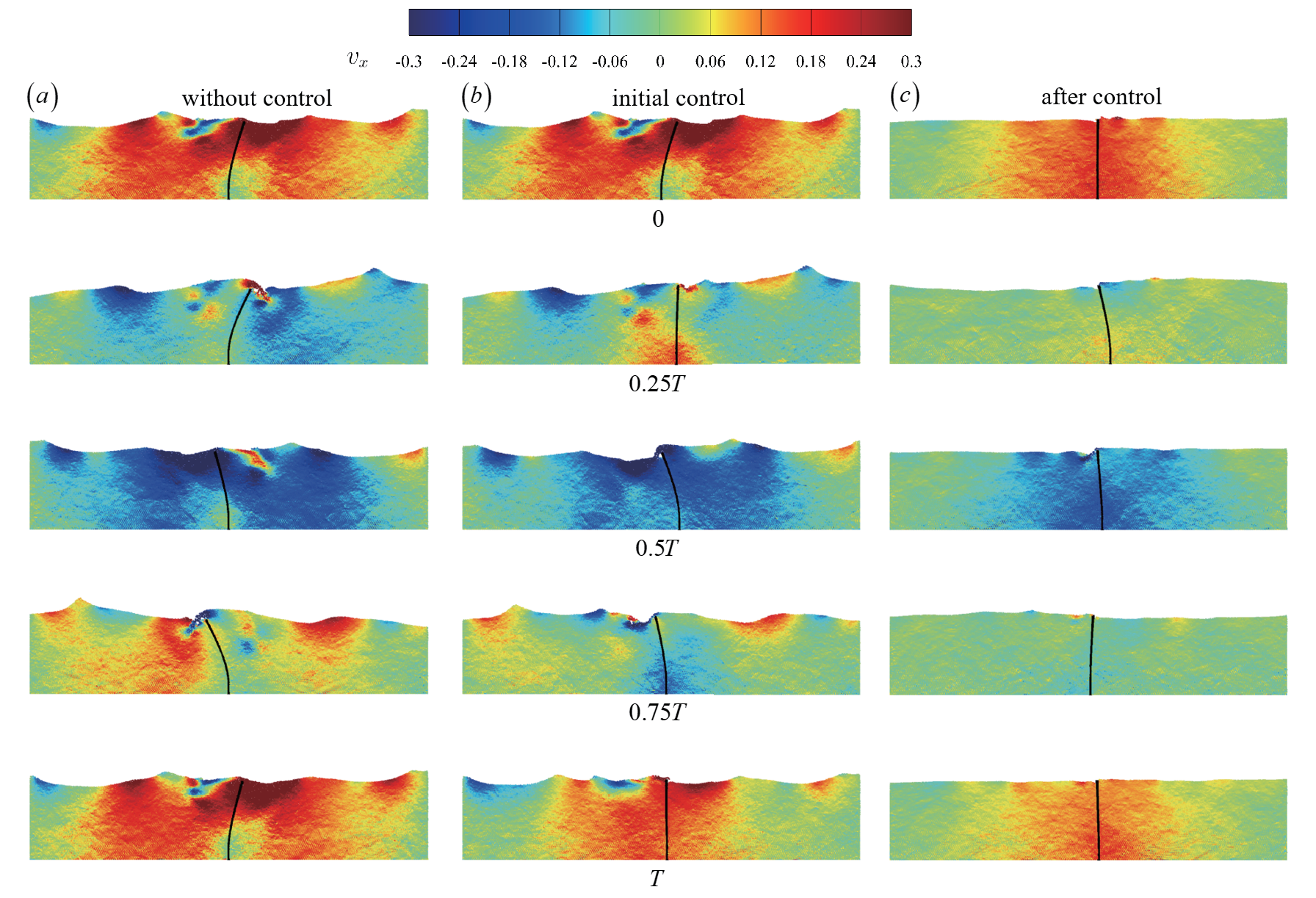}
\caption{Contour plots of tank sloshing within a wave period at different stages: (a) without control, (b) during the first wave period after control activation, and (c) after the system reaches a controlled, steady state.}
\label{fig:6}
\end{figure}

\begin{figure}[htbp]
    \centering
    \begin{subfigure}{\textwidth}
        \centering
        \includegraphics[width=\textwidth]{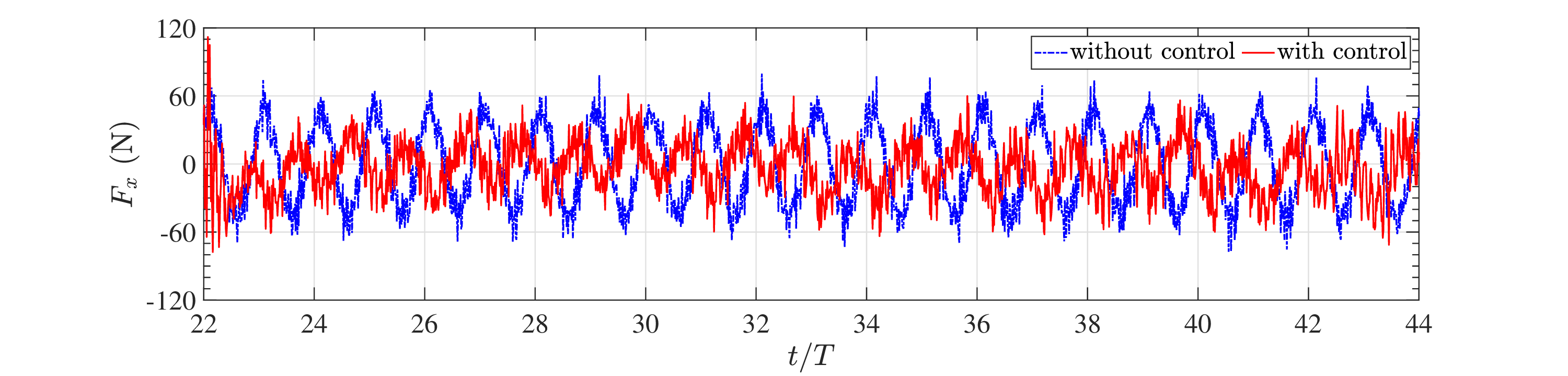}
        \caption{force in $x$-direction}
    \end{subfigure}
    
    \begin{subfigure}{\textwidth}
        \centering
        \includegraphics[width=\textwidth]{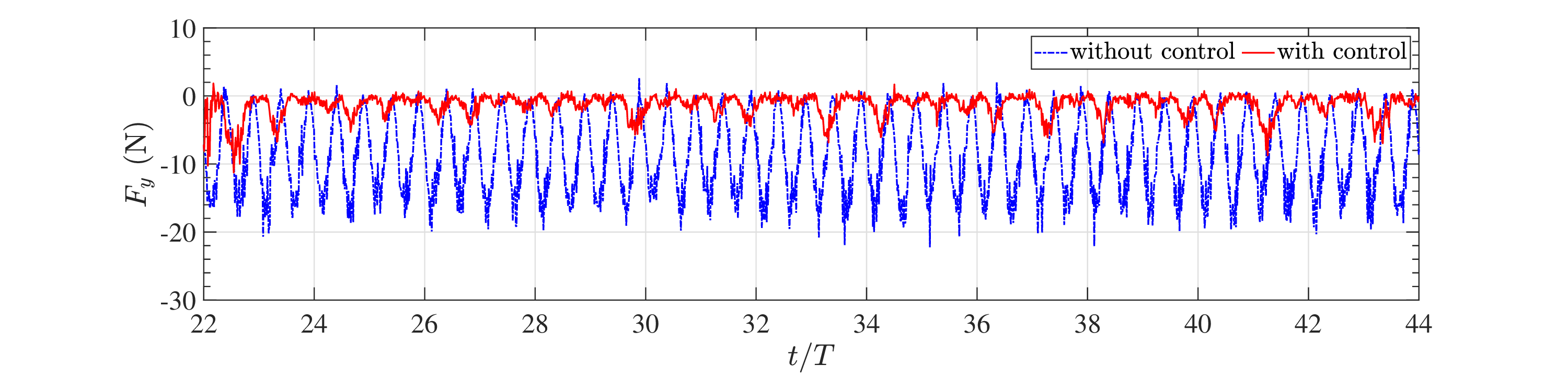}
        \caption{force in $y$-direction}
    \end{subfigure}

    \begin{subfigure}{\textwidth}
        \centering
        \includegraphics[width=\textwidth]{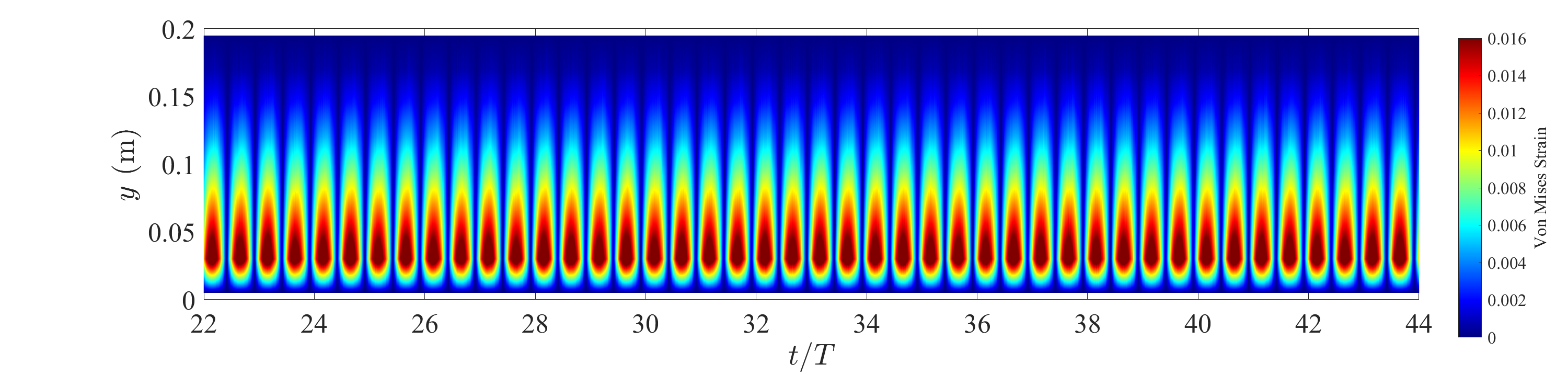}
        \caption{Von Mises strain without control}
    \end{subfigure}
    
    \begin{subfigure}{\textwidth}
        \centering
        \includegraphics[width=\textwidth]{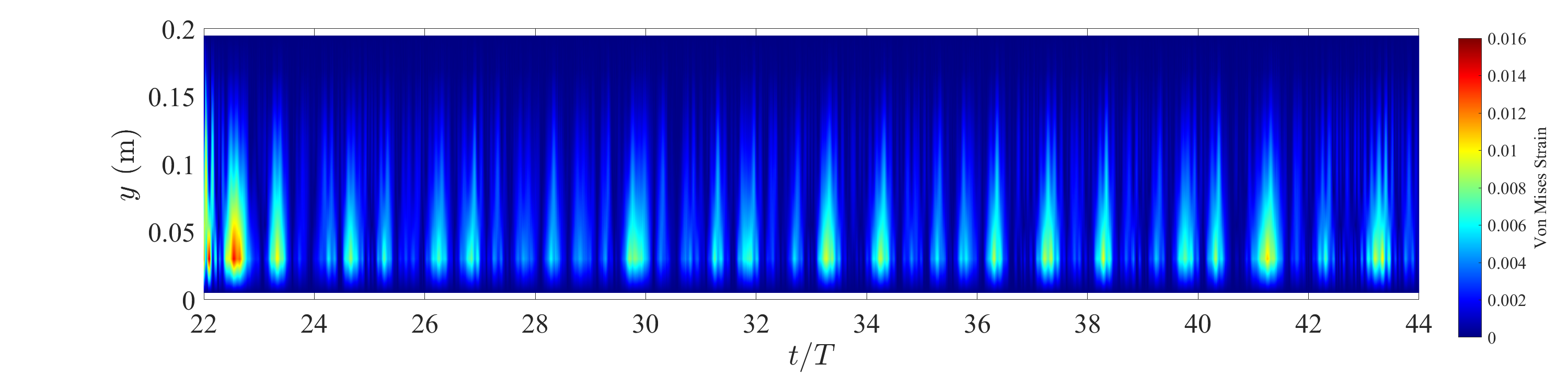}
        \caption{Von Mises strain with control}
    \end{subfigure}

    \caption{Time history of the hydrodynamic force acting on the baffle in the (a) \(x\)-direction and (b) \(y\)-direction, and the corresponding evolution of the Von Mises strain distribution along the baffle height: (c) without control and (d) with control.}
    \label{fig:7}
\end{figure}

\begin{figure}[htbp]
    \centering
    \begin{subfigure}{\textwidth}
        \centering
        \includegraphics[width=\textwidth]{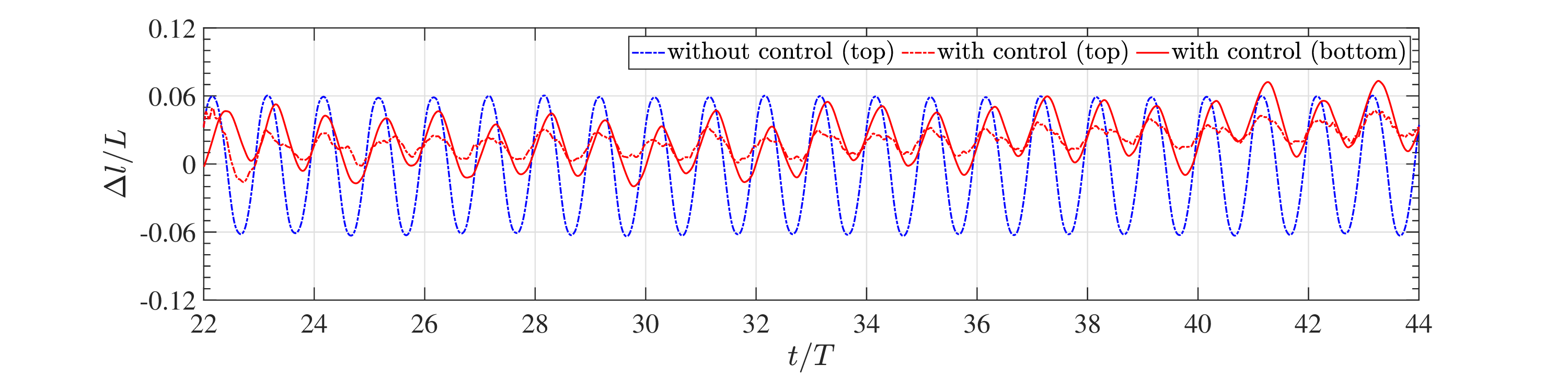}
        \caption{displacement}
    \end{subfigure}
    
    \begin{subfigure}{\textwidth}
        \centering
        \includegraphics[width=\textwidth]{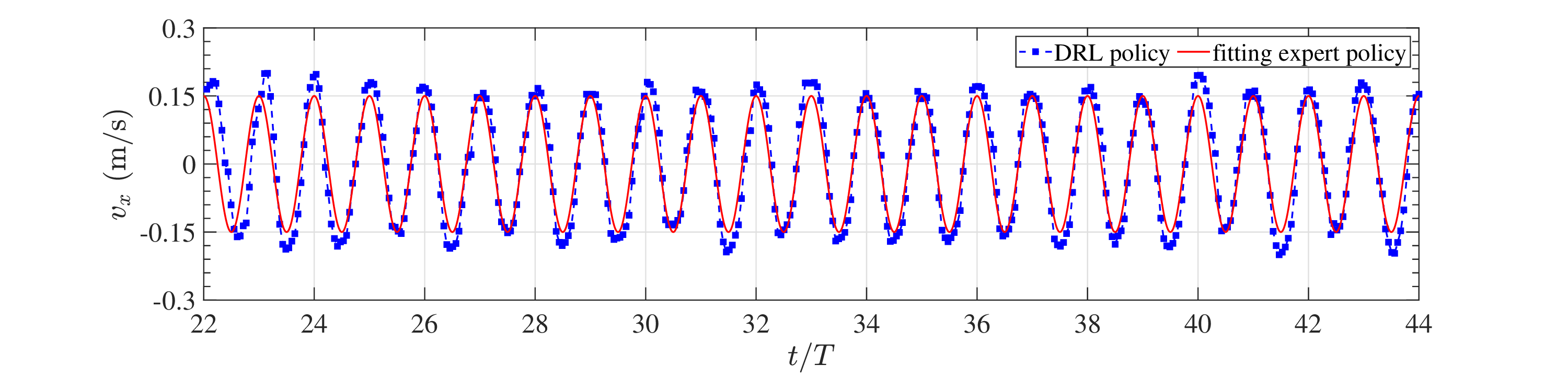}
        \caption{velocity}
    \end{subfigure}

    \caption{Time history of the baffle's (a) displacement trajectory and (b) velocity, which is controlled with DRL policy in $x$-direction.}
    \label{fig:8}
\end{figure}

An analysis of the $x$-direction velocity contours in Fig.~\ref{fig:6} reveals that the motion of the baffle significantly alters the flow field structure. As shown in Fig.~\ref{fig:7} (a), during the initial 0--0.25 $T$ of the wave period without control, the baffle experiences a positive hydrodynamic force in the $x$-direction. This is primarily due to the strong rightward fluid motion in the upper region of the tank ($y > 0.1\,\mathrm{m}$), which exerts substantial thrust on the upper part of the baffle. As a result, the lower half ($y \leq 0.1\,\mathrm{m}$) undergoes notable deformation, as indicated by the Von Mises strain distribution in Fig.~\ref{fig:7} (c). Meanwhile, the lower fluid region remains nearly stationary or exhibits slight leftward flow. Under active control, the baffle moves rightward during this phase (Fig.~\ref{fig:8}), resulting in a leftward (opposing) hydrodynamic force from the fluid. This implies that the baffle performs negative work on the fluid, as evidenced by the force profile in Fig.~\ref{fig:7}. Notably, the vertical ($y$-direction) force is significantly reduced, and the corresponding Von Mises strain decreases markedly, indicating reduced structural deformation. The baffle, therefore, behaves more like a rigid vertical barrier. Due to elastic recovery, the upper part of the baffle also performs negative work on the fluid. Similar behavior is observed during the 0.5--0.75 $T$ interval. In contrast, during 0.25--0.5 $T$ and 0.75--1.0 $T$, both $x$- and $y$-direction forces are relatively small, and the baffle’s motion aligns with the fluid force, indicating that it performs positive work on the fluid during these intervals.

Further insights can be drawn from Fig.~\ref{fig:8}, which shows the baffle's displacement and velocity time histories. The baffle exhibits an apparent periodic motion, with a frequency closely matching the external excitation. However, the center of its oscillation is slightly shifted from the original equilibrium position, introducing asymmetry into the flow field and a sloshing frequency of \( 0.45 \, \mathrm{Hz} \), as mentioned before. Based on this learned DRL policy, we propose a fitted expert velocity policy that captures this periodic behavior. The policy is formulated as a cosine function and can be expressed as:
\begin{equation}
u_x^b = A_b \cos(\omega_b t),
\label{eq:19}
\end{equation}
where $A_b = 0.15 \, \mathrm{m/s}$ is the oscillation amplitude got from the training data, $\omega_b = \omega_e$. Based on this formulation, the corresponding displacement of the baffle can be derived as $X_b = A_b /\omega_e \sin(\omega_e t)$, indicating that under the expert policy, the center of baffle oscillation aligns with the center of the tank, and remains in phase with the external excitation. To further validate the effectiveness of this policy, it is applied within both 2D and 3D numerical simulation environments.

\begin{figure}[htbp]
    \centering

    \begin{subfigure}{\textwidth}
        \centering
        \includegraphics[width=\textwidth]{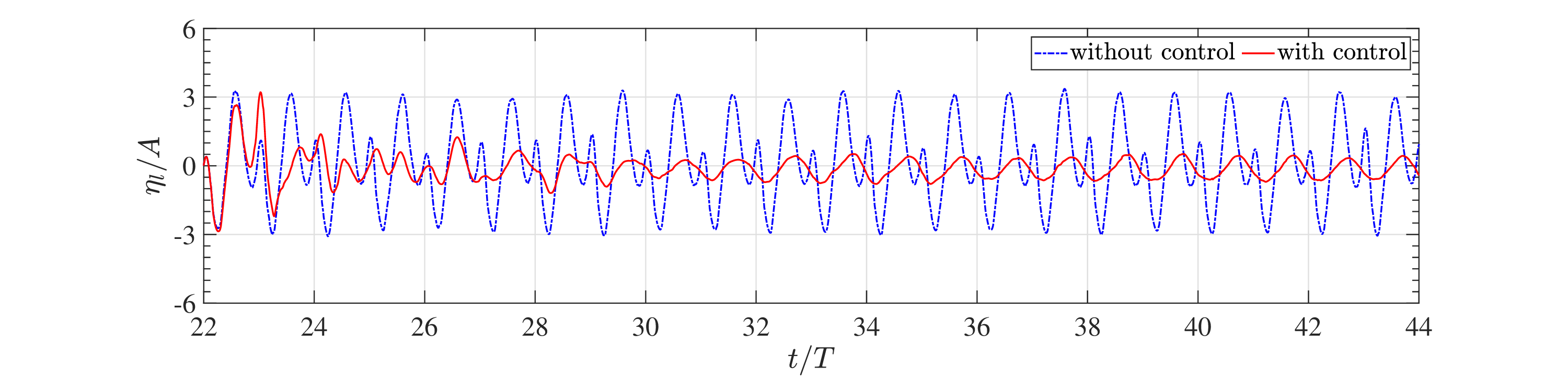}
        \caption{free surface height}
    \end{subfigure}
    
    \begin{subfigure}{\textwidth}
        \centering
        \includegraphics[width=\textwidth]{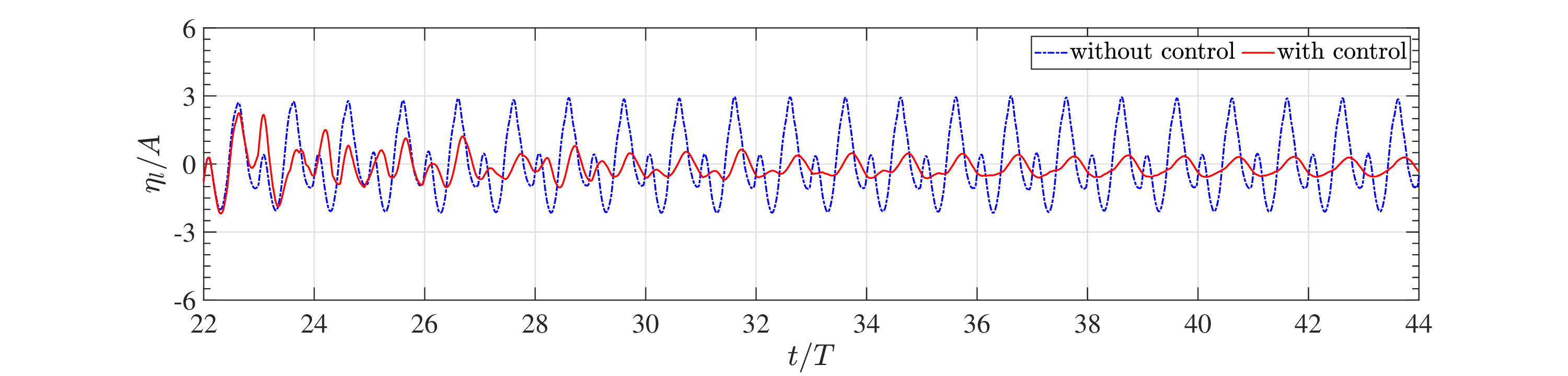}
        \caption{free surface height}
    \end{subfigure}

    \begin{subfigure}{\textwidth}
        \centering
        \includegraphics[width=\textwidth]{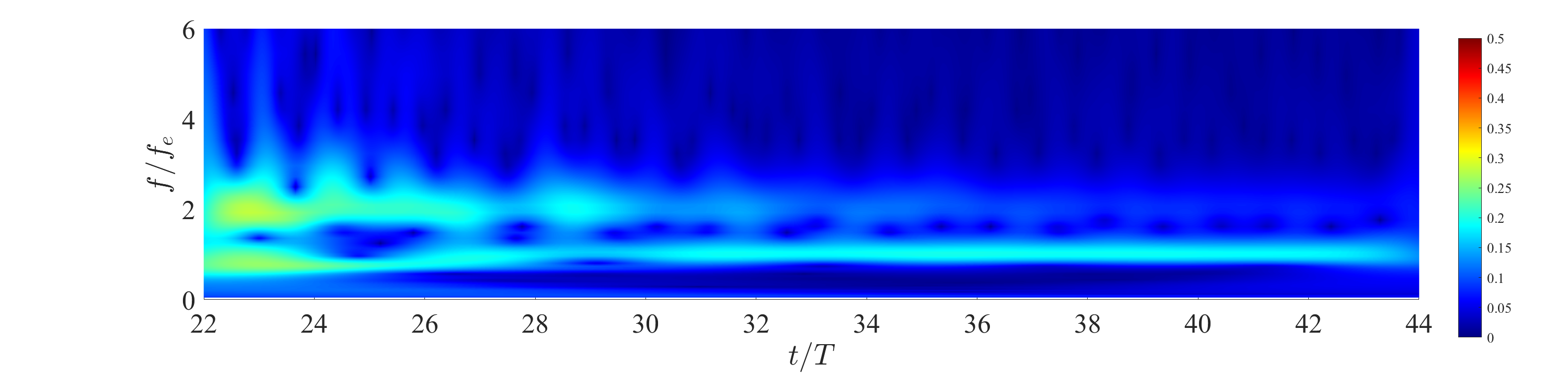}
        \caption{wavelet transform}
    \end{subfigure}

    \caption{Comparison of the time history of free surface height (a) 2D, (b) 3D, and (c) corresponding wavelet transforms at the left wall with expert policy in 3D.}
    \label{fig:9}
\end{figure}

\begin{figure}
\centering
\includegraphics[width=\textwidth]{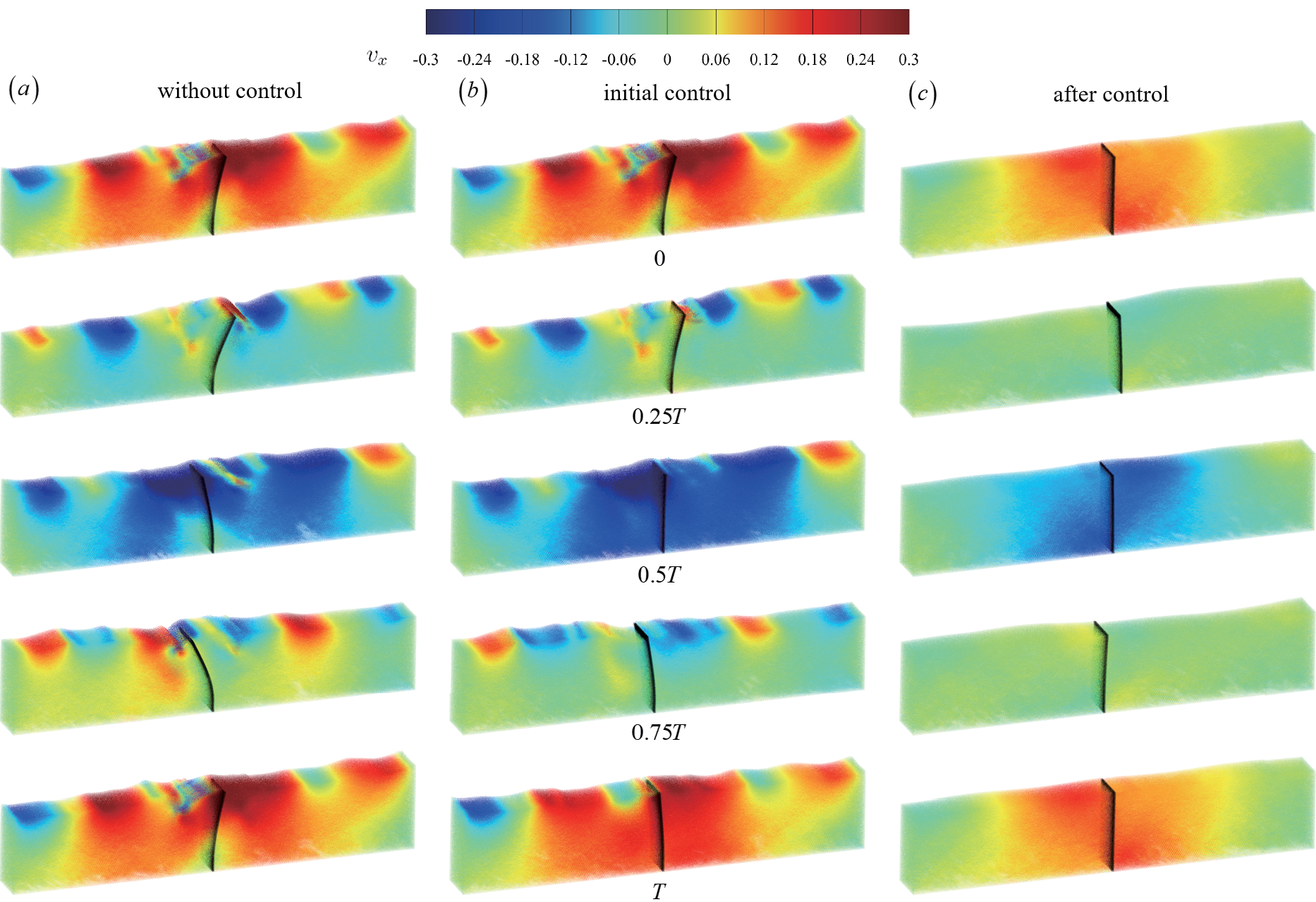}
\caption{Contour plots of tank sloshing within a wave period at different stages in 3D: (a) without control, (b) during the first wave period after control activation, and (c) after the system reaches a controlled, steady state.}
\label{fig:10}
\end{figure}

\begin{figure}[htbp]
    \centering
    \begin{subfigure}{\textwidth}
        \centering
        \includegraphics[width=\textwidth]{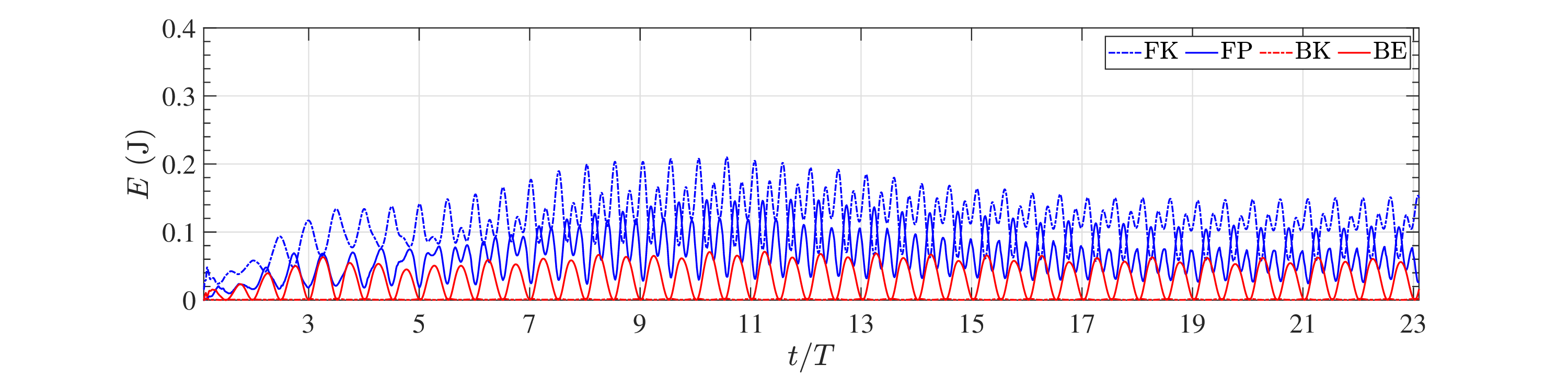}
        \caption{baffle motion only}
    \end{subfigure}
    
    \begin{subfigure}{\textwidth}
        \centering
        \includegraphics[width=\textwidth]{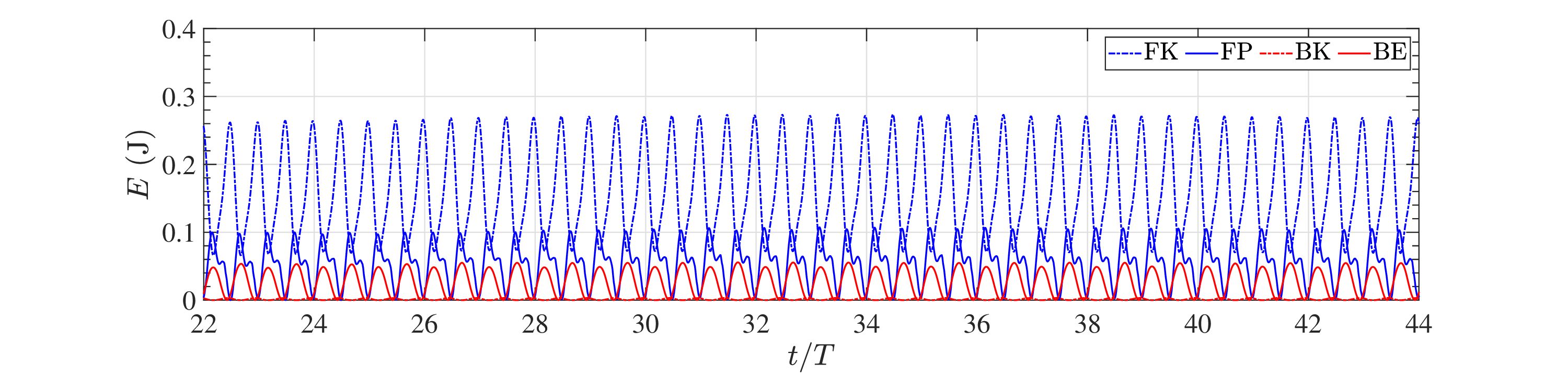}
        \caption{without control}
    \end{subfigure}

    \begin{subfigure}{\textwidth}
        \centering
        \includegraphics[width=\textwidth]{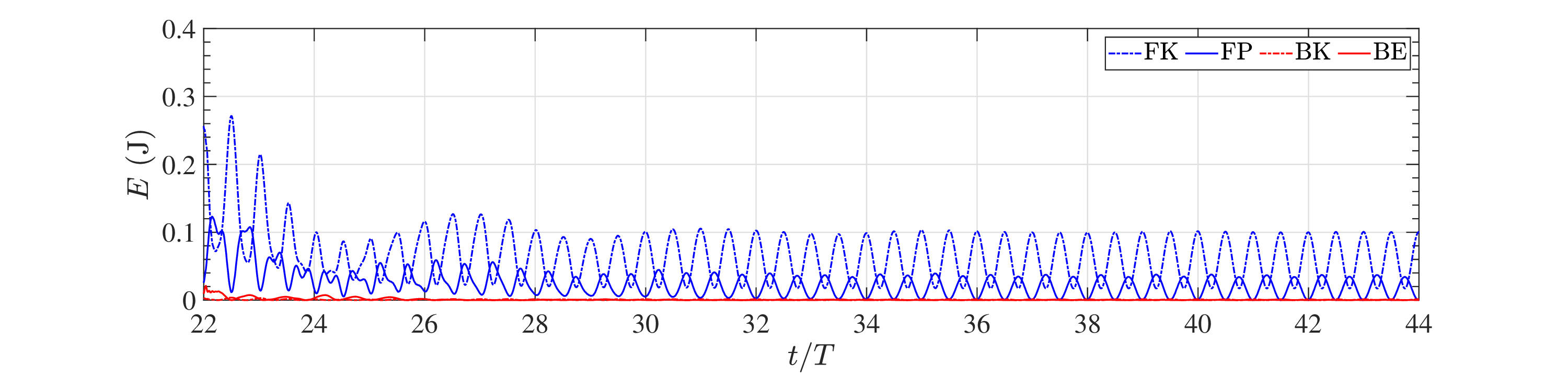}
        \caption{with control}
    \end{subfigure}

    \caption{Time histories of fluid kinetic energy (FK), fluid potential energy (FP), baffle kinetic energy (BK) and baffle elastic energy (BE): (a) without tank sloshing (baffle motion only), (b) without control, and (c) with active control. }
    \label{fig:11}
\end{figure}

As shown in Fig.~\ref{fig:9} (a) and (b), the time history of the free surface height at the midplane ($z = 0$) confirms that the expert control policy effectively suppresses sloshing, achieving an overall amplitude reduction of approximately 76.9\% in both 2D and 3D cases. However, the suppression is less pronounced compared to the 2D case with the DRL policy. This discrepancy may be attributed to the control policy enforcing strictly periodic motion centered at the geometric midpoint of the tank. As a result, even after stabilization, the free surface exhibits sustained periodic oscillations. This is further supported by the wavelet analysis in Fig.~\ref{fig:9} (c), which shows that the dominant frequency remains close to the natural frequency of the tank–baffle system, around \(1.1\,\mathrm{Hz}\). Operating near resonance can lead to a slight amplification of the sloshing response. Moreover, the influence of 3D effects in the $z$-direction under the current condition appears negligible, as indicated by the velocity contours shown in Fig.~\ref{fig:10}. In addition, the elastic energy stored in the baffle is calculated as:
\begin{equation}
\label{eq:20}
B_{e} = \sum_i \frac{1}{2} \, \text{tr} \left( \mathbb{P}_i \cdot \mathbb{E}_i \right) V_i.
\end{equation}
Then, the evolution of energy conversion within the tank–baffle system is illustrated in Fig.~\ref{fig:11}. It can be observed that, in the case where only the baffle undergoes motion without sloshing, the system gains energy significantly due to the positive work done by the baffle. The resulting energy level is comparable in magnitude and order to that generated by the sloshing motion. In contrast, when active control is applied, the fluid's kinetic and gravitational potential energy are substantially reduced. Notably, the elastic energy of the baffle remains close to zero throughout the controlled motion, indicating that the baffle behaves approximately as a rigid baffle.

\subsection{Effects of control strategies}
\begin{figure}
\centering
\includegraphics[width=\textwidth]{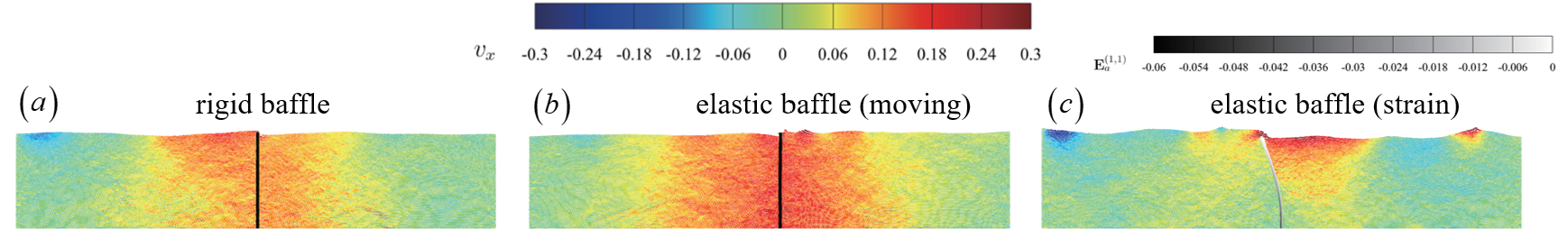}
\caption{Contour plots of tank sloshing under different control strategies in the steady state: (a) moving rigid baffle, (b) moving elastic baffle, and (c) elastic baffle with active strain.}
\label{fig:12}
\end{figure}

Considering that the baffle exhibited negligible deformation after stabilization during active control in the previous section, this section investigates the performance differences among various control strategies under the same external excitation and tank conditions (\( h = 0.2 \, \mathrm{m}, f_e = 1.1 \, \mathrm{Hz}\) ). Specifically, we add two approaches: moving a rigid baffle and applying active strain to induce elastic bending, as shown in Fig.~\ref{fig:12}. Once the control reaches a steady state, we can see that the flow fields around the moving rigid baffle and the moving elastic baffle become comparable at the phase $t/T=0$. In both cases, the velocity distribution on either side of the baffle remains relatively smooth and continuous. In contrast, when control is applied via active strain, a velocity discontinuity emerges across the baffle interface, indicating stronger localized flow disturbances. 

\begin{figure}[htbp]
    \centering
    \begin{subfigure}{\textwidth}
        \centering
        \includegraphics[width=\textwidth]{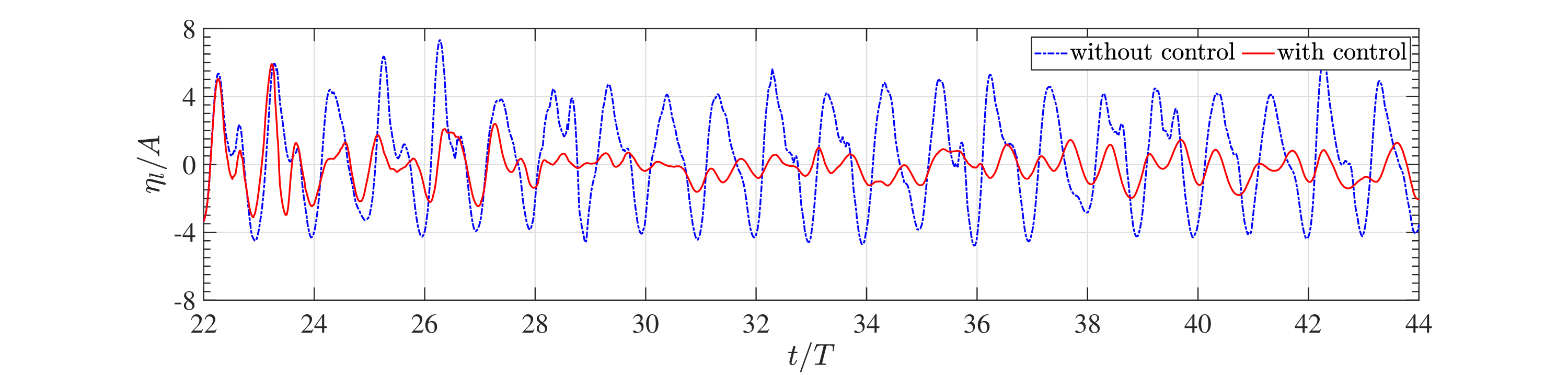}
        \caption{rigid baffle}
    \end{subfigure}
    
    \begin{subfigure}{\textwidth}
        \centering
        \includegraphics[width=\textwidth]{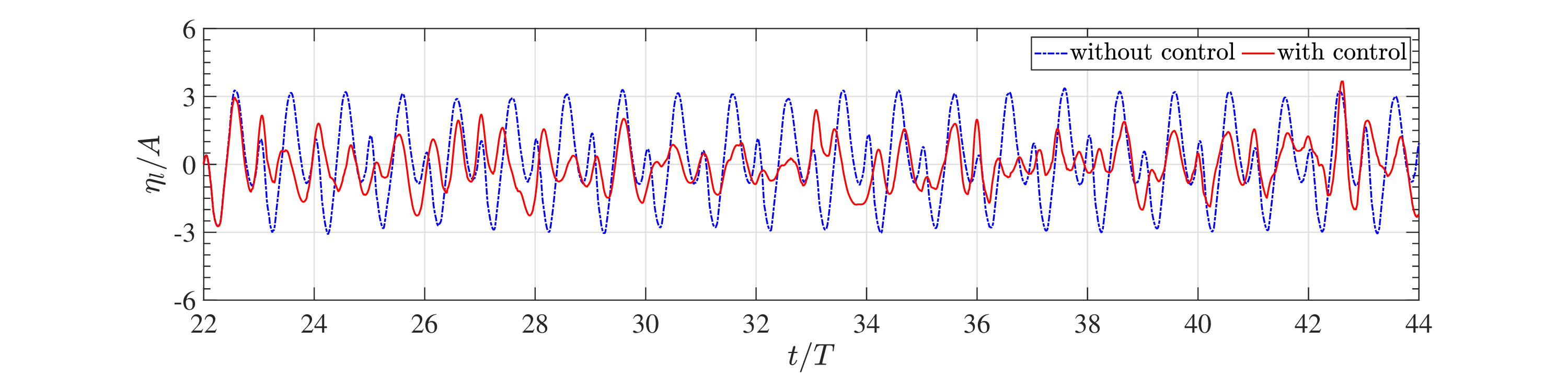}
        \caption{elastic baffle}
    \end{subfigure}

    \caption{Comparison of the time history of free surface height at the left wall: (a) rigid baffle, (b) elastic baffle.}
    \label{fig:13}
\end{figure}

\begin{figure}[htbp]
    \centering
    \begin{subfigure}{\textwidth}
        \centering
        \includegraphics[width=\textwidth]{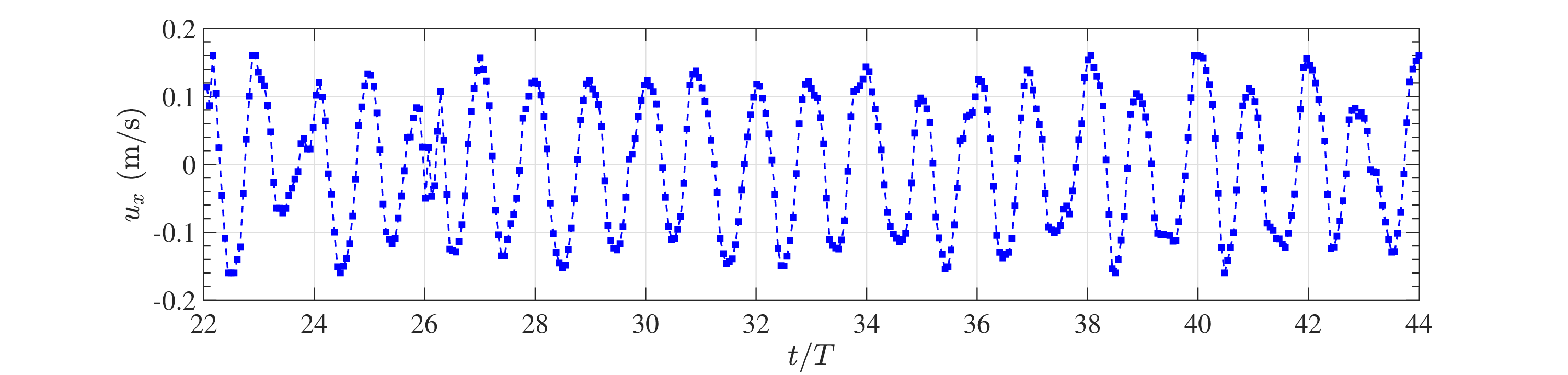}
        \caption{rigid baffle}
    \end{subfigure}
    
    \begin{subfigure}{\textwidth}
        \centering
        \includegraphics[width=\textwidth]{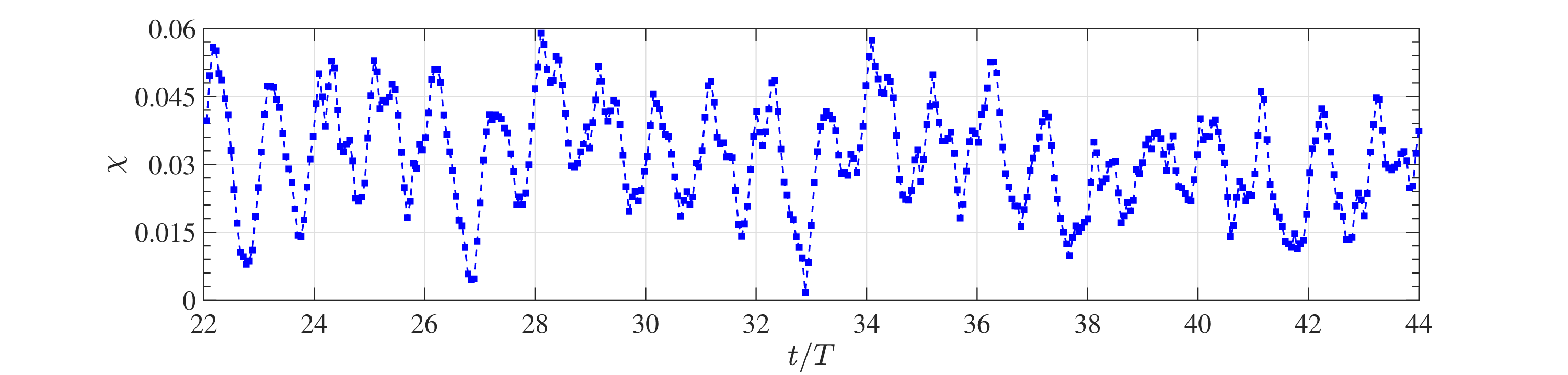}
        \caption{elastic baffle}
    \end{subfigure}

    \caption{Time history of the baffle’s (a) velocity and (b) active strain, which are controlled by the DRL policy.}
    \label{fig:14}
\end{figure}

Figure~\ref{fig:13} further illustrates the free surface height before and after control is applied. In the uncontrolled cases, the oscillation amplitude is higher for the rigid baffle than the elastic one. However, when lateral motion is applied to the rigid baffle, sloshing is effectively suppressed by 73.02\%, reducing $\eta_{mean}$ to \(0.0068\, \mathrm{m}\), though it still exceeds that achieved by the moving elastic baffle (\(0.0027\, \mathrm{m}\)). On the other hand, active strain control reduces sloshing by only 41.5\%, suggesting a less effective suppression mechanism. Besides, from Fig.~\ref{fig:14}, we can see that the velocity profile of the rigid baffle exhibits a clear periodic pattern similar to that of the moving elastic baffle. However, noticeable variations in amplitude are observed across different wave cycles. In contrast, the imposed strain fluctuates more dramatically and lacks strong periodicity, introducing additional nonlinear effects into the system. This makes it more difficult for the system to stabilize, weakening the overall sloshing suppression performance.

\begin{figure}[htbp]
    \centering
    \begin{subfigure}{0.48\textwidth}
        \centering
        \includegraphics[width=\linewidth]{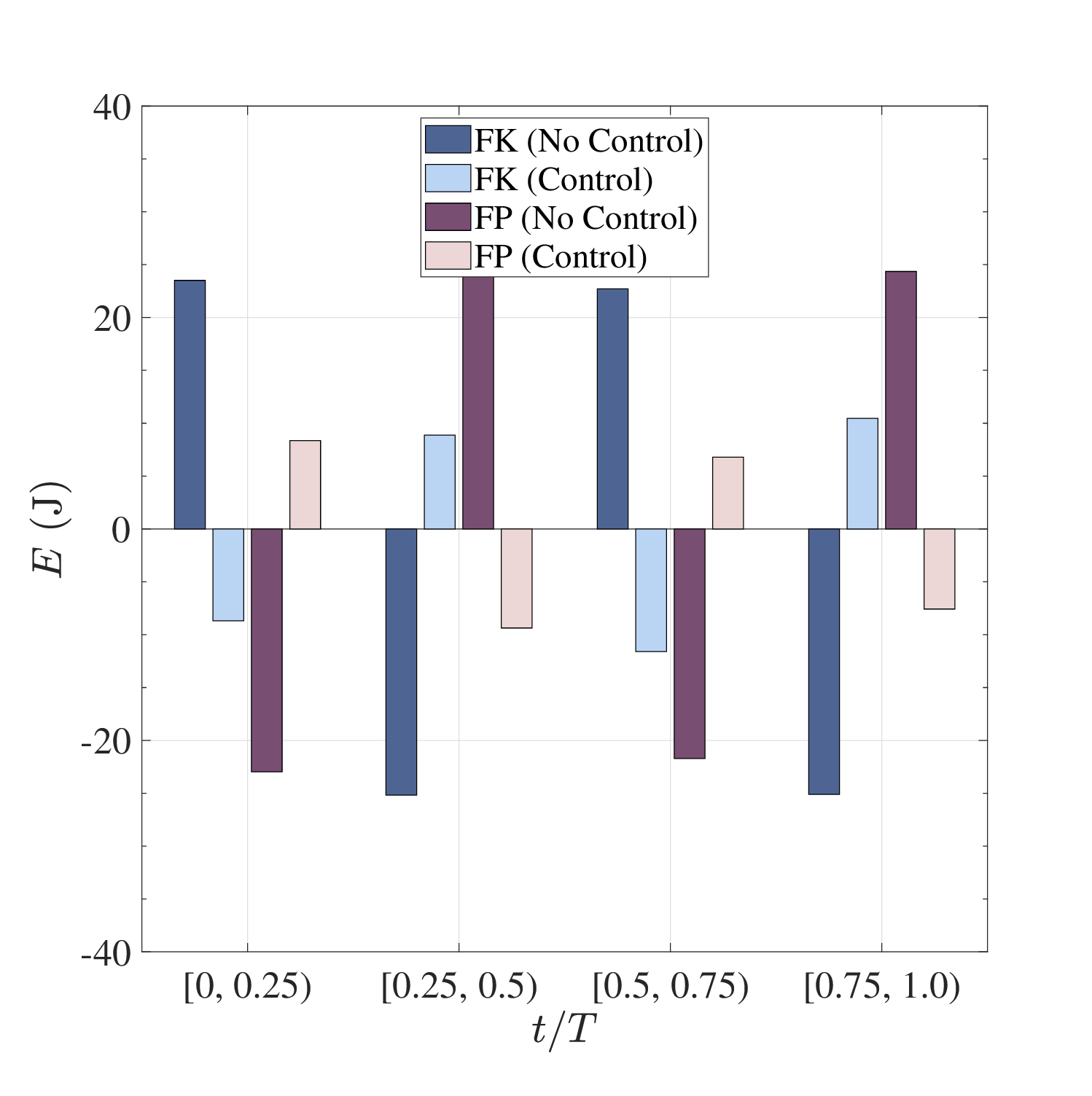}
        \caption{rigid baffle}
    \end{subfigure}
    \hfill
    \begin{subfigure}{0.48\textwidth}
        \centering
        \includegraphics[width=\linewidth]{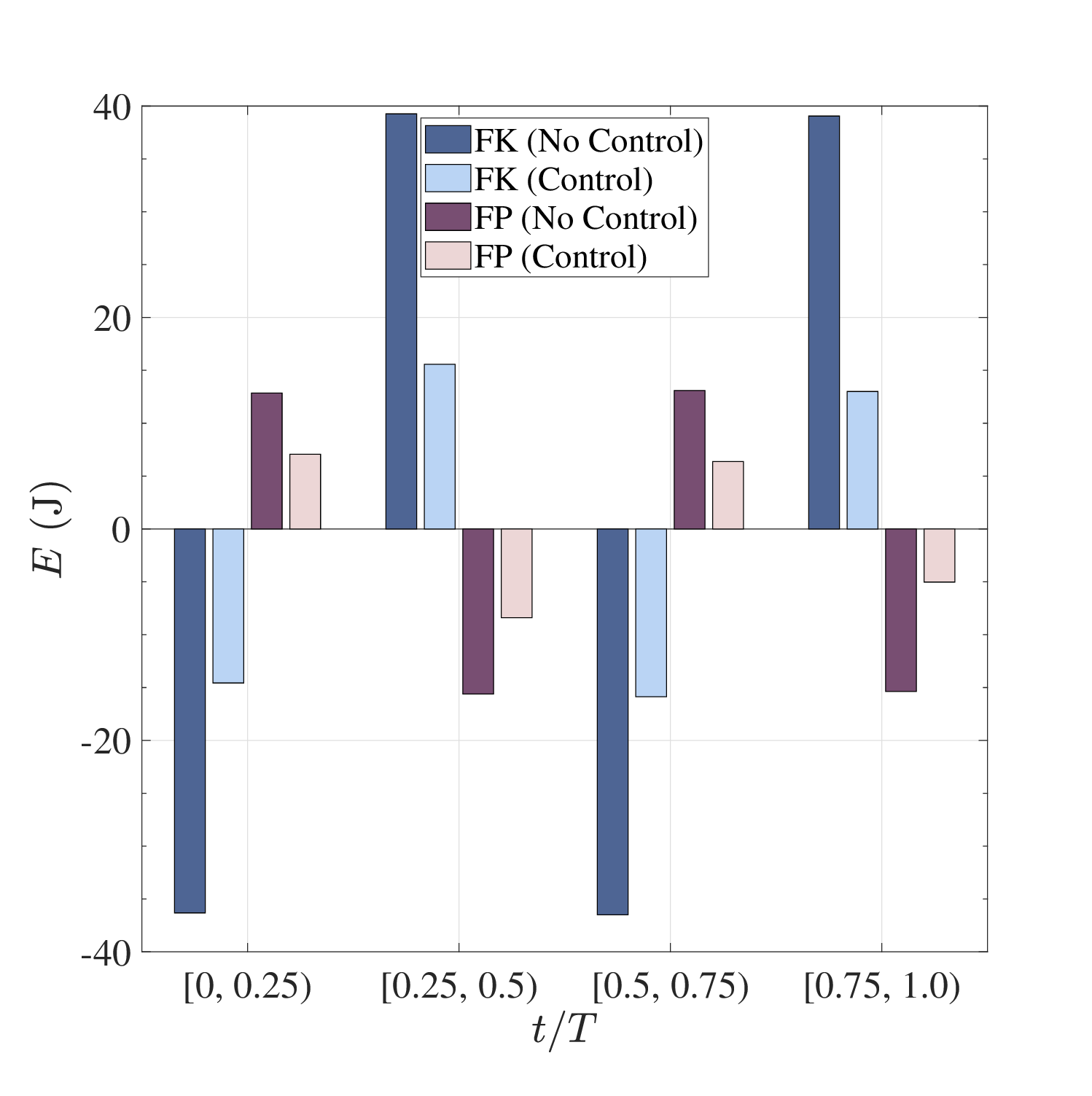}
        \caption{elastic baffle (strain)}
    \end{subfigure}

    \caption{Cumulative variation of fluid kinetic and potential energy in each quarter-period interval over 20 sloshing cycles. Negative values represent the amount of energy reduction in the interval.}
    \label{fig:15}
\end{figure}

\begin{figure}
\centering
\includegraphics[width=0.85\textwidth]{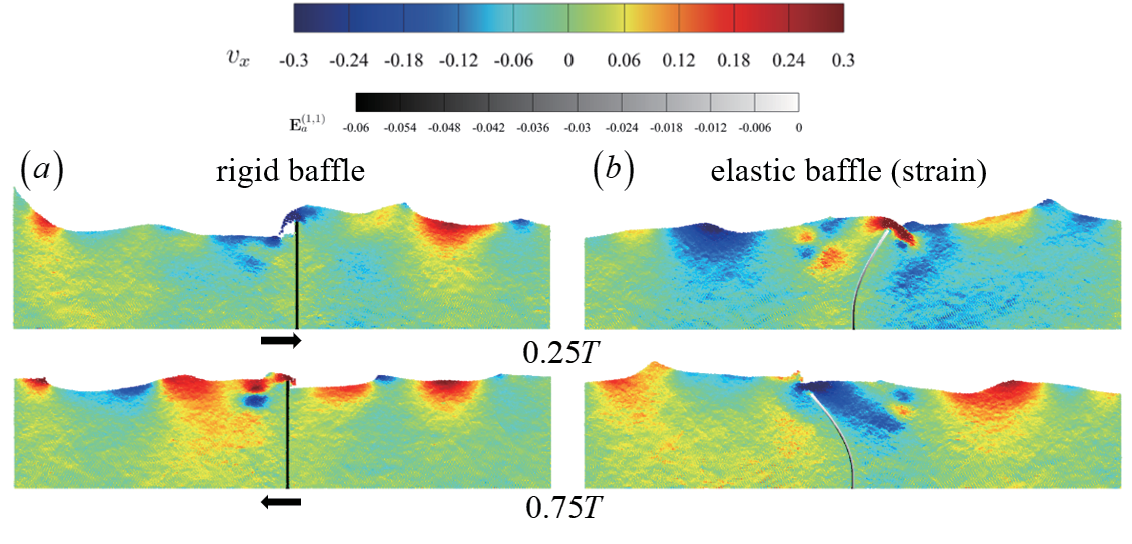}
\caption{Contour plots of tank sloshing within the first wave period after control activation: (a) rigid baffle, (b) elastic baffle.}
\label{fig:16}
\end{figure}

Under the given excitation condition, the displacement of the rigid baffle leads to a net reduction of \(2.75\,\mathrm{J}\) in the total fluid energy. As shown in Fig.~\ref{fig:15}, the introduction of baffle motion effectively reverses the natural energy oscillation cycle. Specifically, in the interval 0--0.25 $T$, where kinetic energy would typically increase while potential energy decreases, the control strategy instead causes kinetic energy to drop and potential energy to rise. Overall, during 0--0.5 $T$, the baffle performs negative work on the fluid, though the magnitude is moderate (approximately \(0.82\,\mathrm{J}\)). In the interval 0.5--0.75 $T$, more significant negative work is observed, resulting in a total energy reduction of \(4.8\,\mathrm{J}\). Conversely, in 0.75--1.0 $T$, the baffle performs positive work, increasing the energy by \(2.88\,\mathrm{J}\). The velocity contours in Fig.~\ref{fig:16} further reveal that during 0--0.25 $T$ and 0.5--0.75 $T$, the direction of baffle motion (left and right) opposes the flow direction induced by the external excitation. This leads to fluid compression and velocity attenuation in front of the baffle, thereby increasing the potential energy. In contrast, when using active strain to deform the elastic baffle, although a total energy reduction of \(1.85\,\mathrm{J}\) is achieved, the periodic characteristic of the energy oscillation remains largely unchanged. The suppression mainly occurs in the 0--0.25 $T$ and 0.5--0.75 $T$ intervals. For instance, in the first quarter-cycle, the applied strain induces a rightward bending of the baffle from its mid-to-lower section. Compared with the uncontrolled case, the deformation enhances the displacement toward the right. At this time, leftward-flowing water near the baffle must overcome the deformation, resulting in a net energy loss.

Overall, for the considered scenario, both rigid and elastic baffle movement strategies outperform active strain control in suppressing sloshing.

\subsection{Effects of external excitations over a broad frequency range}
\begin{figure}
\centering
\includegraphics[width=0.85\textwidth]{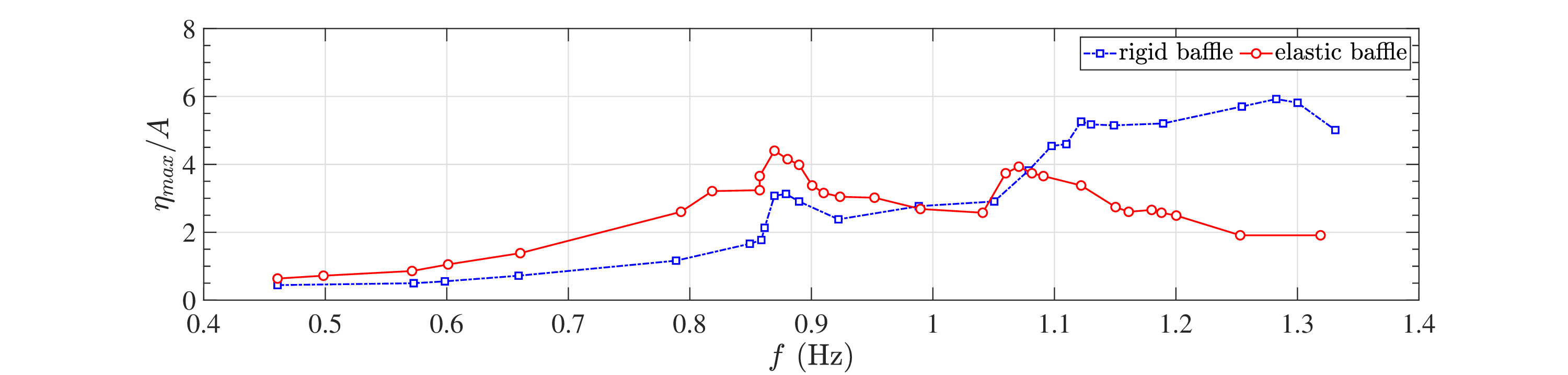}
\caption{Maximum free surface height on the wall for rigid and elastic baffles with fixed bases at different excitation frequencies.}
\label{fig:17}
\end{figure}

Based on the experimental observations in Fig.~\ref{fig:17} \cite{ren2023experimental}, when the external excitation frequency \( f_e \) is below \(0.8\,\mathrm{Hz}\), the sloshing amplitude remains limited. For the rigid baffle, sloshing becomes significantly amplified at \( f_e = 0.9\,\mathrm{Hz} \), which is close to the sub-resonance frequency, approximately half of the second natural frequency (\( f_2 \approx 1.76\,\mathrm{Hz} \)) of a tank-baffle system with water depth \( h = 0.2\,\mathrm{m} \) and tank length \( L = 0.5\,\mathrm{m} \). For the elastic baffle around \( f_e = 0.9\,\mathrm{Hz} \), it is affected by its wet natural frequency (\( f_{\mathrm{wet}} \approx 0.93\,\mathrm{Hz} \)), in addition to the sub-resonance effects. As a result, its suppression performance is inferior to that of the rigid baffle. At \( f_e = 1.1\,\mathrm{Hz} \), both rigid and elastic baffles are near resonance with their respective first natural frequencies (\( f_1^{\mathrm{rigid}} \approx 1.12\,\mathrm{Hz} \), \( f_1^{\mathrm{elastic}} \approx 1.05\,\mathrm{Hz} \)), reducing the control effectiveness. When the excitation frequency exceeds \(1.1\,\mathrm{Hz}\), particularly around \(1.3\,\mathrm{Hz}\), the suppression performance of the rigid baffle declines. This can be attributed to the excitation frequency approaching twice the first natural frequency of the tank without a baffle, thereby exciting higher-mode resonance. In contrast, the elastic baffle maintains better suppression under high-frequency excitation. Its deformability alters the system's symmetry and shifts the natural frequency, reducing resonance amplification.

Two representative frequencies are considered in this section to evaluate the performance of the DRL-based control policy for rigid and elastic baffles: \( f_e = 0.9\,\mathrm{Hz} \) and \( f_e = 1.3\,\mathrm{Hz} \).

\begin{figure}[htbp]
    \centering
    \begin{subfigure}{\textwidth}
        \centering
        \includegraphics[width=\textwidth]{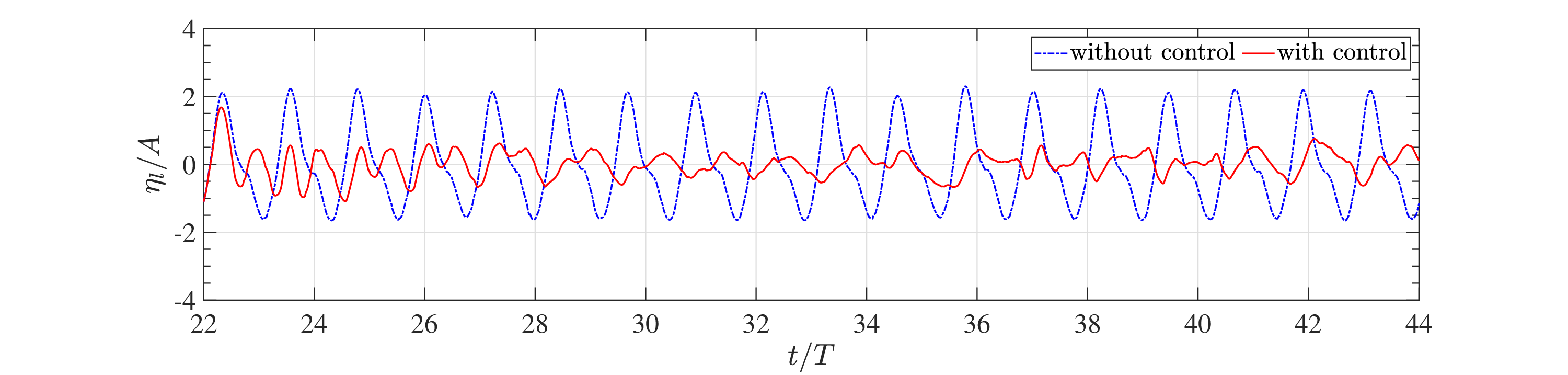}
        \caption{rigid baffle, \(f_e = 0.9 \, \mathrm{Hz}\)}
    \end{subfigure}
    
    \begin{subfigure}{\textwidth}
        \centering
        \includegraphics[width=\textwidth]{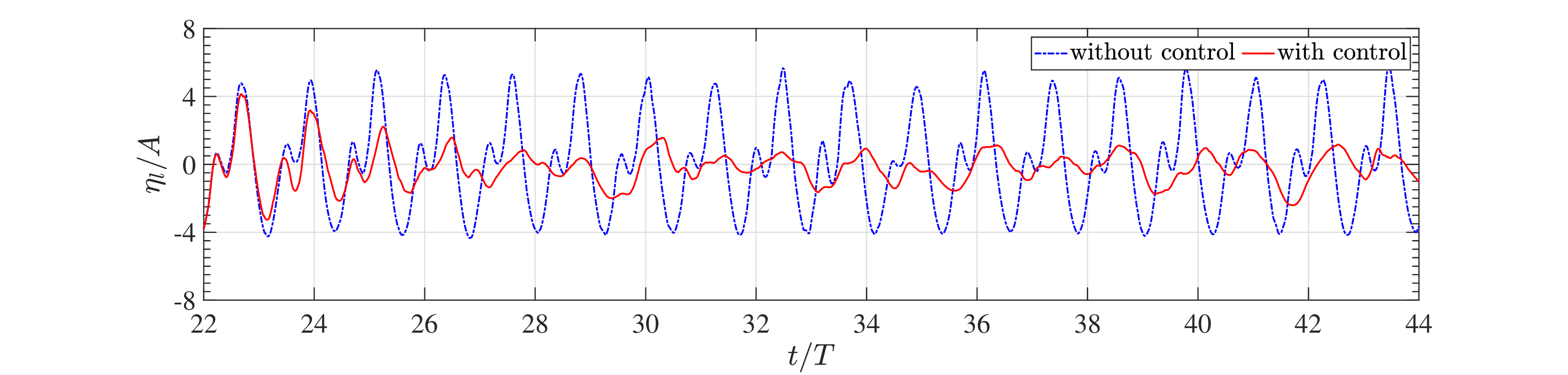}
        \caption{elastic baffle, \(f_e = 0.9 \, \mathrm{Hz}\)}
    \end{subfigure}

    \begin{subfigure}{\textwidth}
        \centering
        \includegraphics[width=\textwidth]{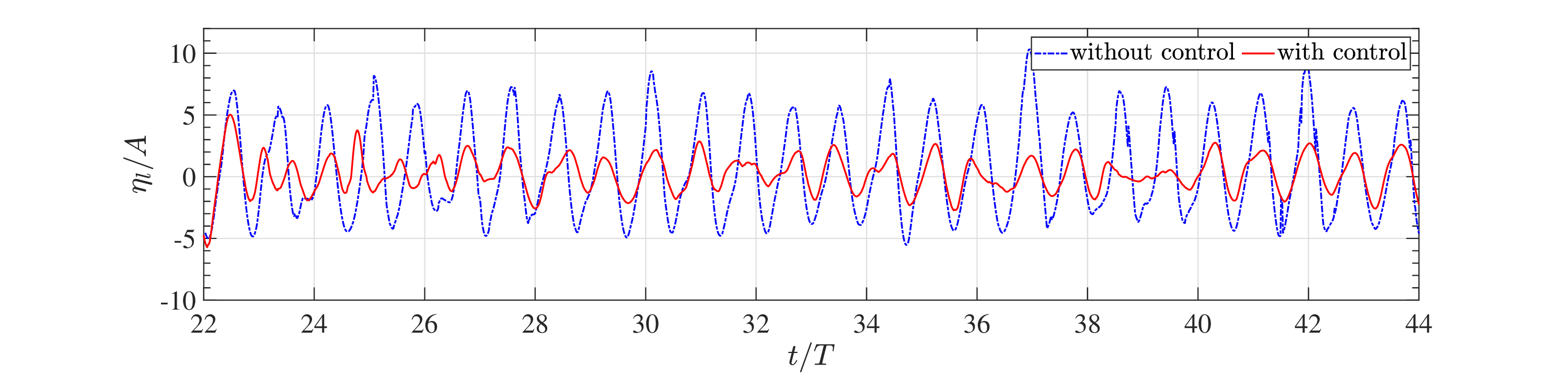}
        \caption{rigid baffle, \(f_e = 1.3 \, \mathrm{Hz}\)}
    \end{subfigure}

    \begin{subfigure}{\textwidth}
        \centering
        \includegraphics[width=\textwidth]{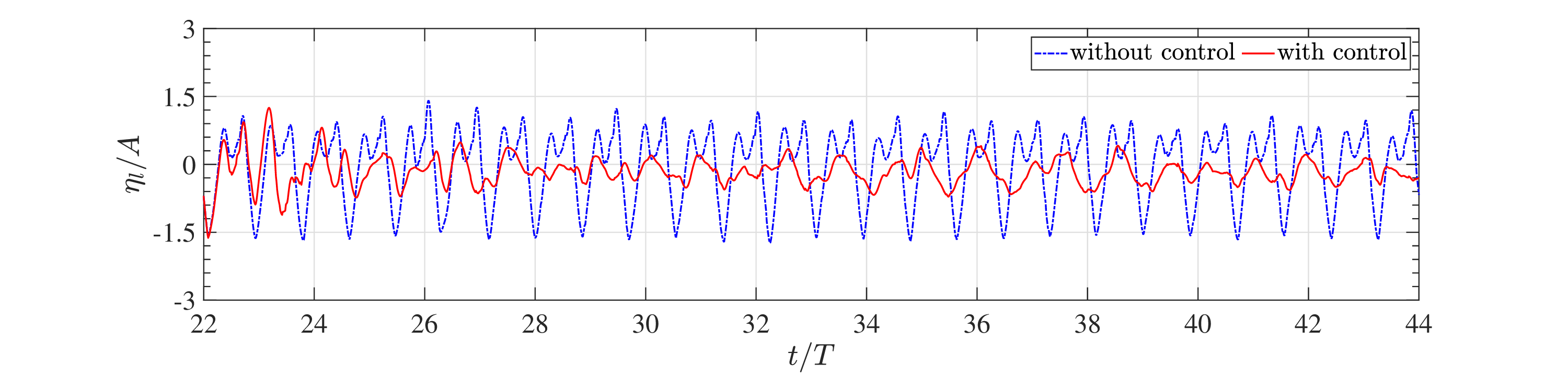}
        \caption{rigid baffle, \(f_e = 1.3 \, \mathrm{Hz}\)}
    \end{subfigure}

    \caption{Comparison of the time history of free surface height at the left wall under different excitation frequencies: (a) (c) rigid baffle, (b) (d) elastic baffle.}
    \label{fig:18}
\end{figure}

As shown in Fig.~\ref{fig:18}, actively controlling the baffle's motion reduces the free surface height under both excitation frequencies. In addition to the fundamental frequency (e.g., \(0.9\,\mathrm{Hz}\)), higher-order harmonics such as \(1.8\,\mathrm{Hz}\) and \(2.7\,\mathrm{Hz}\) are observed due to strong nonlinear FSI. Active baffle motion effectively suppresses these harmonic components, thereby enhancing sloshing mitigation. Specifically, for \( f_e = 0.9\,\mathrm{Hz} \), the mean free surface amplitude $\eta_{mean}$ over 10 stabilized sloshing periods is reduced from \( 0.010\,\mathrm{m} \) to \( 0.0028\,\mathrm{m} \) for the rigid baffle, and from \( 0.0224\,\mathrm{m} \) to \( 0.007\,\mathrm{m} \) for the elastic baffle. At \( f_e = 1.3\,\mathrm{Hz} \), the amplitude decreases from \( 0.033\,\mathrm{m} \) to \( 0.0116\,\mathrm{m} \) for the rigid baffle, and from \( 0.007\,\mathrm{m} \) to \( 0.002\,\mathrm{m} \) for the elastic baffle. The results summarized in Table~\ref{table:1} show that the elastic baffle exhibits superior performance at higher frequencies (\(1.1\) and \(1.3\,\mathrm{Hz}\)), while the rigid baffle performs better under lower-frequency excitation (\(0.9\,\mathrm{Hz}\)). This trend is consistent with the passive sloshing behavior observed without active control.

\begin{figure}[htbp]
    \centering
    \begin{subfigure}{0.48\textwidth}
        \centering
        \includegraphics[width=\linewidth]{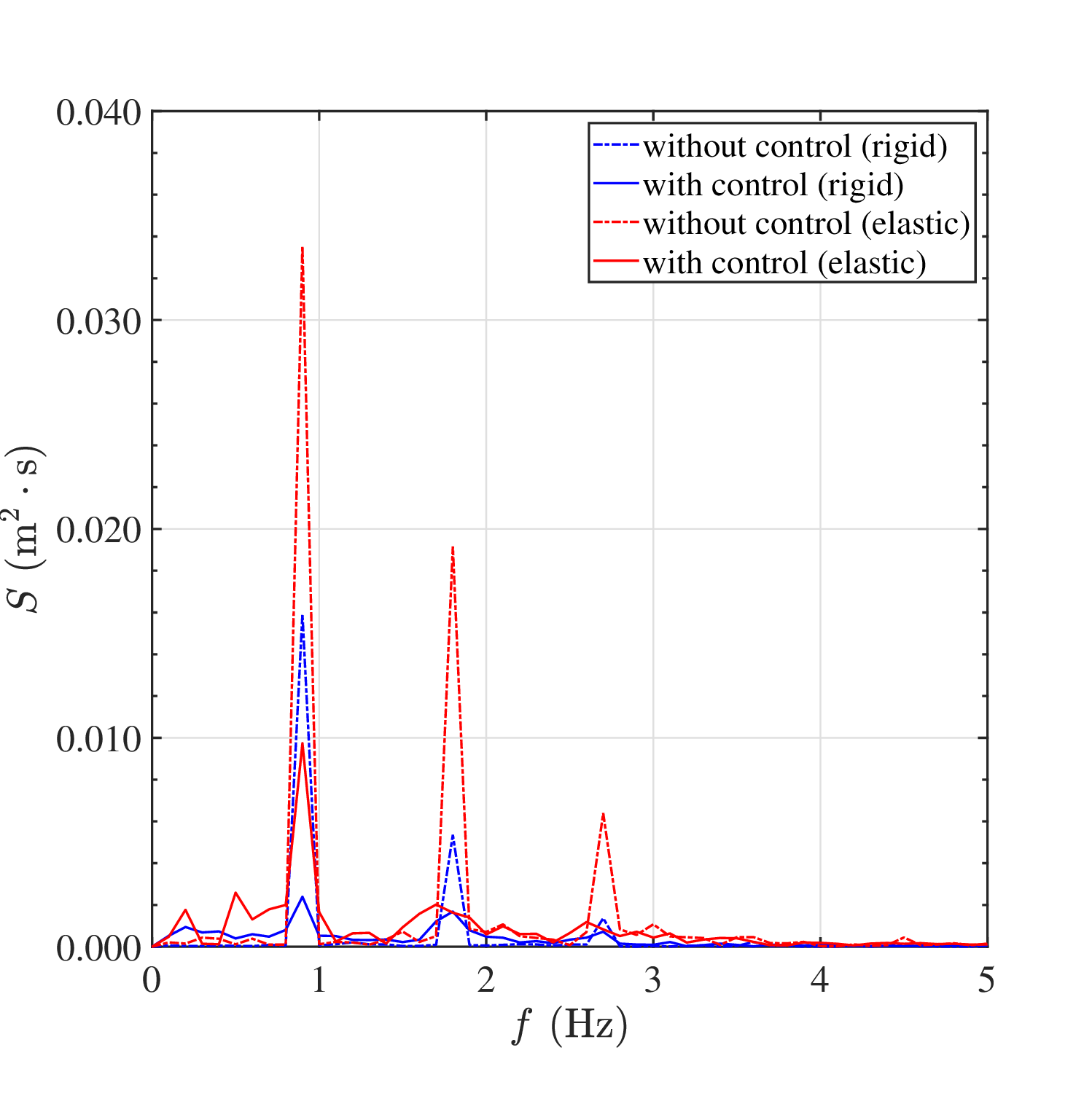}
        \caption{\(0.9 \, \mathrm{Hz}\)}
    \end{subfigure}
    \hfill
    \begin{subfigure}{0.48\textwidth}
        \centering
        \includegraphics[width=\linewidth]{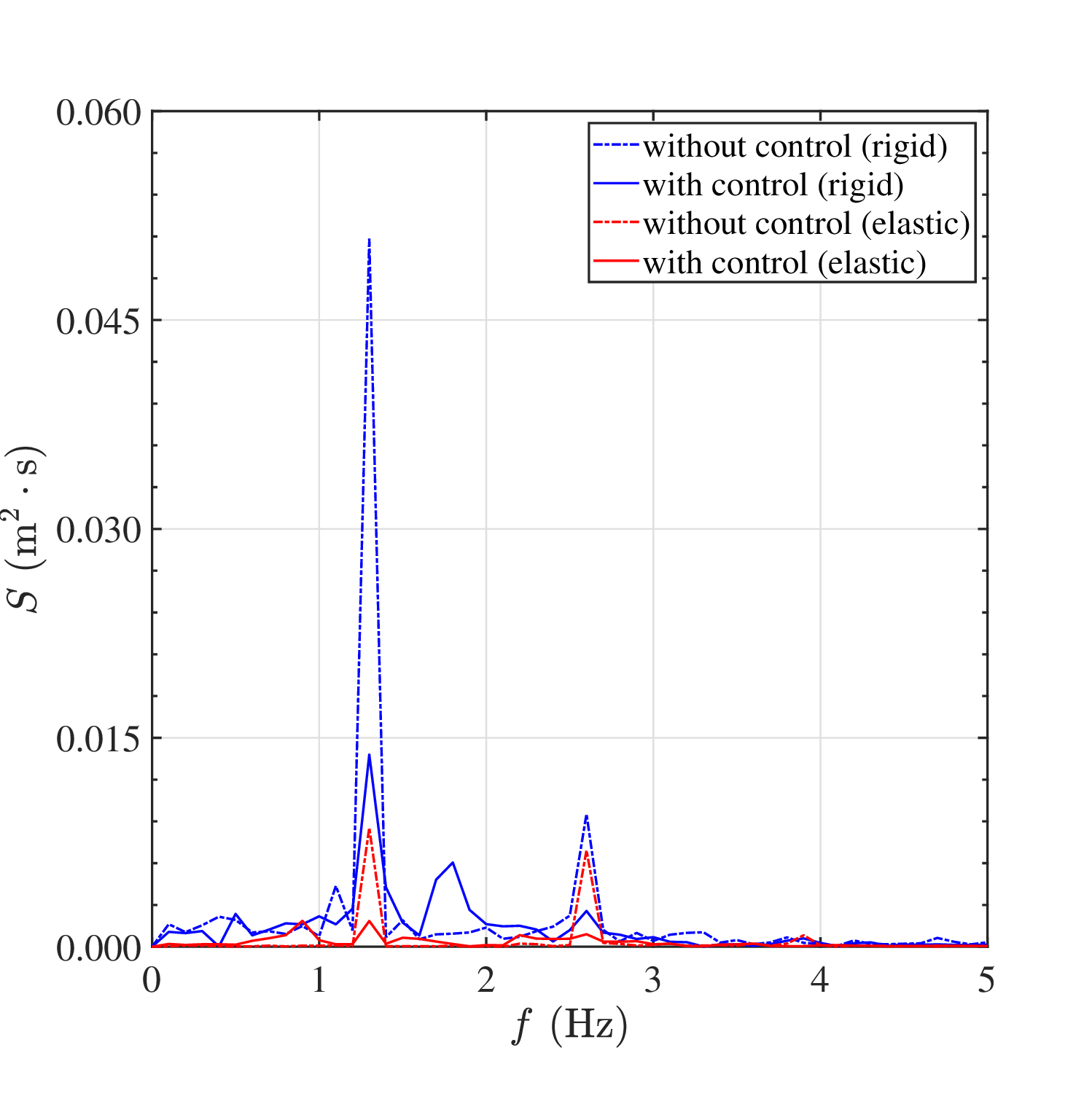}
        \caption{\(1.3 \, \mathrm{Hz}\)}
    \end{subfigure}

    \caption{Fast Fourier Transform (FFT) of the time history of free surface height at the left wall under different excitation frequencies: (a) \(0.9 \, \mathrm{Hz}\), (b) \(1.3 \, \mathrm{Hz}\).}
    \label{fig:19}
\end{figure}

\begin{table}[htbp]
    \centering
    \caption{Mean free surface amplitude $\eta_{mean}$ before and after control at different excitation frequencies. The values in parentheses represent the percentage reduction.}
    \label{table:1}
    \begin{tabular}{@{}lccc@{}}
        \toprule
        Frequency ($f_e$, Hz) & 0.9 & 1.1 & 1.3 \\ \midrule
        Rigid & 0.01 & 0.025 & 0.033 \\
        (with control) & 0.0028 (72.0\%) & 0.0068 (73.0\%) & 0.0116 (64.8\%) \\ \addlinespace
        Elastic & 0.0224 & 0.0147 & 0.007 \\
        (with control) & 0.007 (68.8\%) & 0.0027 (81.6\%) & 0.002 (71.4\%) \\
        \bottomrule
    \end{tabular}
\end{table}

\begin{figure}[htbp]
    \centering
    \begin{subfigure}{\textwidth}
        \centering
        \includegraphics[width=\textwidth]{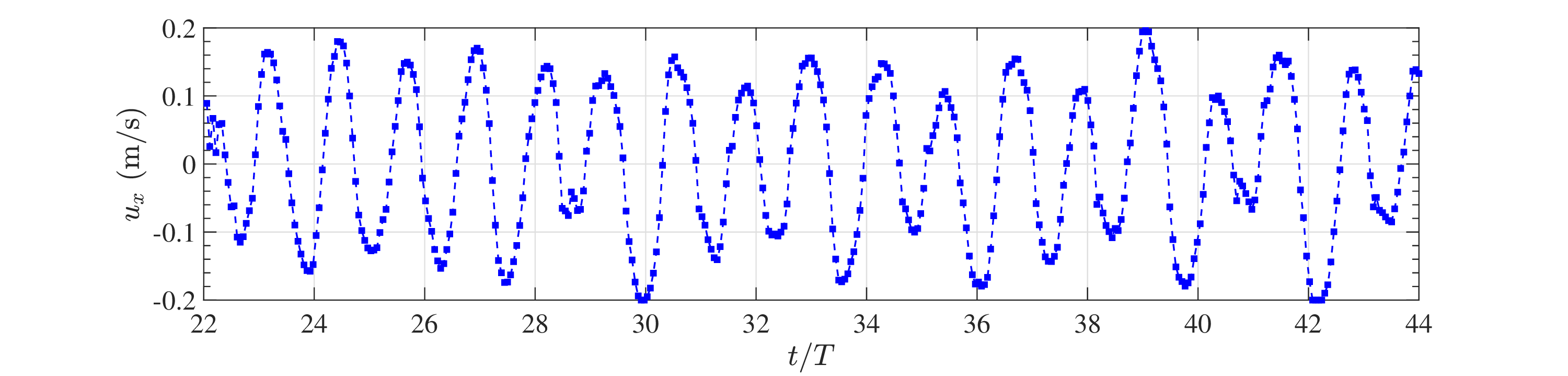}
        \caption{\(0.9 \, \mathrm{Hz}\)}
    \end{subfigure}
    
    \begin{subfigure}{\textwidth}
        \centering
        \includegraphics[width=\textwidth]{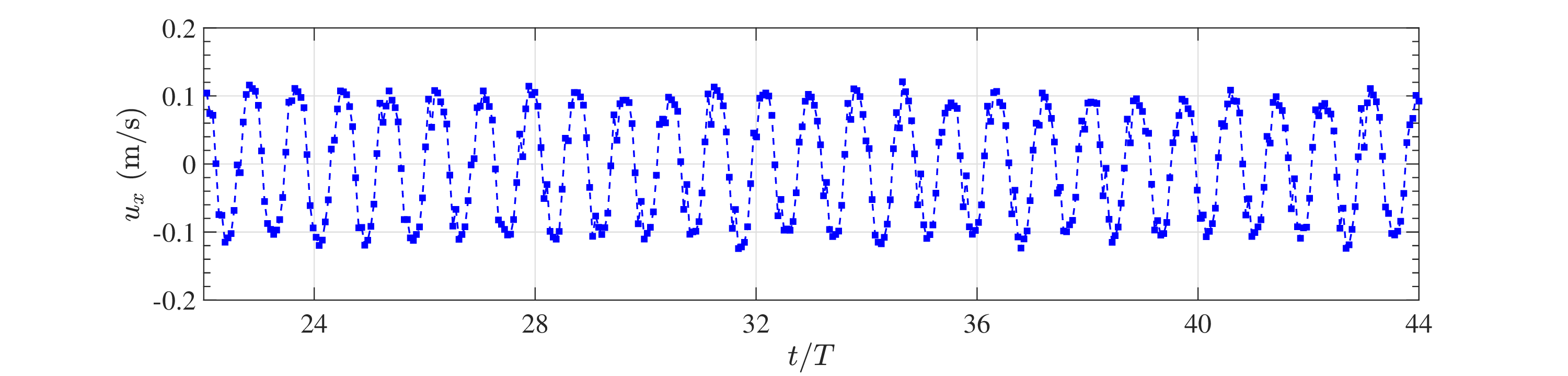}
        \caption{\(1.3 \, \mathrm{Hz}\)}
    \end{subfigure}

    \caption{Time history of the baffle’s velocity at different excitation frequencies: (a) \(0.9 \, \mathrm{Hz}\), (b) \(1.3 \, \mathrm{Hz}\).}
    \label{fig:20}
\end{figure}

\begin{figure}[htbp]
    \centering
    \begin{subfigure}{\textwidth}
        \centering
        \includegraphics[width=\textwidth]{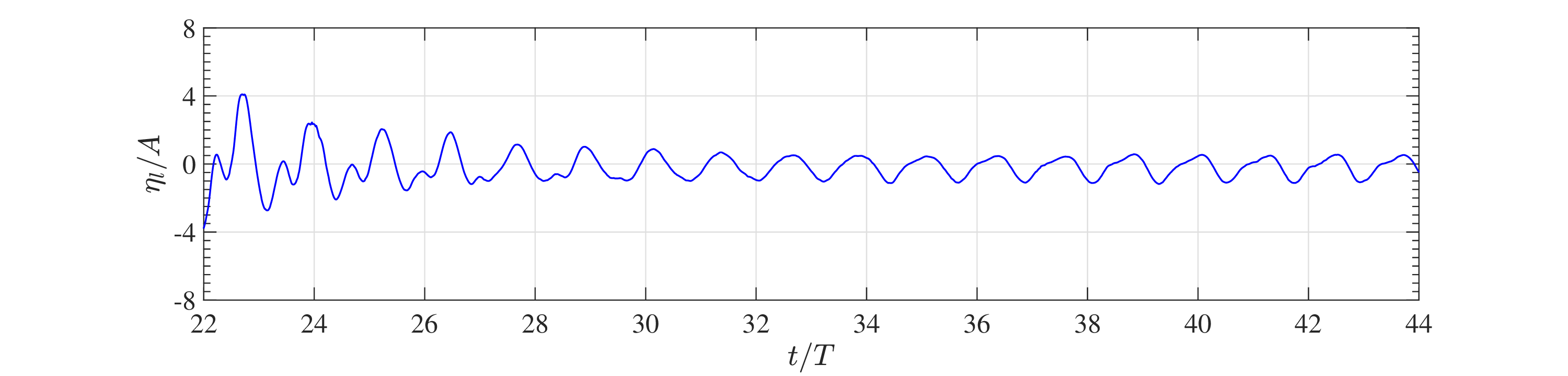}
        \caption{\(0.9 \, \mathrm{Hz}\)}
    \end{subfigure}
    
    \begin{subfigure}{\textwidth}
        \centering
        \includegraphics[width=\textwidth]{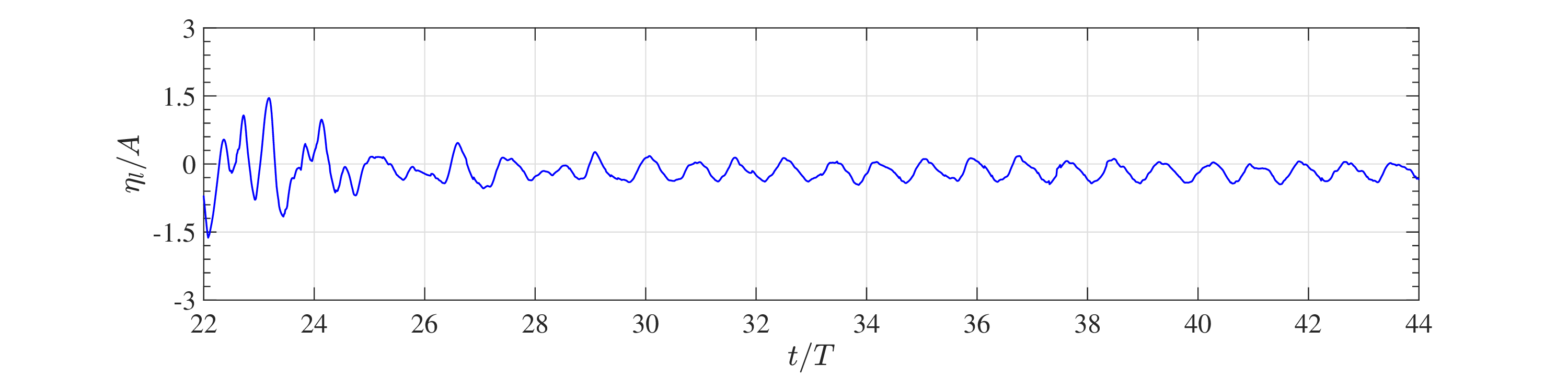}
        \caption{\(1.3 \, \mathrm{Hz}\)}
    \end{subfigure}

    \caption{Time history of the free surface height at the left wall at different excitation frequencies with fitting expert policies: (a) \(0.9 \, \mathrm{Hz}\), (b) \(1.3 \, \mathrm{Hz}\).}
    \label{fig:21}
\end{figure}

Furthermore, as shown in Fig.~\ref{fig:20}, the elastic baffle consistently exhibits periodic motion synchronized with the external excitation frequency across various test cases. As the sloshing intensity increases, the fluid motion becomes more complex, causing the stronger nonlinear characteristics of FSI. This complexity leads to discrepancies in the velocity amplitude prediction when using DRL-trained control policies. Then, the cosine-based expert policies described by Eq.~\ref{eq:19} were fitted to the DRL-learned control policies. For excitation frequencies of \( f_e = 0.9\,\mathrm{Hz} \) and \( f_e = 1.3\,\mathrm{Hz} \), fitting amplitudes of \( A_b = 0.16 \, \mathrm{m/s} \) and \( A_b = 0.12 \, \mathrm{m/s} \) were respectively adopted. The amplitude decreases as the sloshing intensity is reduced. These fitted expert policies were implemented in subsequent simulations. The results in Fig.~\ref{fig:21} demonstrate that the free surface height near the tank wall is significantly suppressed. Due to the periodic nature of baffle motion, the free surface eventually stabilizes into a regular oscillatory pattern after the transient effects subside. Specifically, under \( f_e = 0.9\,\mathrm{Hz} \), the average free surface height \( \eta_{\mathrm{mean}} \) is reduced from \( 0.007\,\mathrm{m} \) to \( 0.0047\,\mathrm{m} \), corresponding to a 79.02\% reduction. For \( f_e = 1.3\,\mathrm{Hz} \), \( \eta_{\mathrm{mean}} \) drops from \( 0.007\,\mathrm{m} \) to \( 0.0014\,\mathrm{m} \), achieving an 80.0\% reduction.

\subsection{Effects of water depths}
Previous studies~\cite{ren2023experimental, ren2023numerical} have shown that when the water depth does not exceed the height of the baffle, the first natural frequency of the tank–baffle system remains nearly unchanged. It is approximately equal to a tank with the same water depth and half-width without any baffles. However, the first natural frequency drops significantly once the water depth exceeds the baffle height. In this section, two additional scenarios with water depths of \( h = 0.15\,\mathrm{m} \) and \( h = 0.25\,\mathrm{m} \) are examined to facilitate comparative analysis, where the first natural frequencies of the tank–baffle system are approximately \( 1.07\,\mathrm{Hz} \) and \( 0.595\,\mathrm{Hz} \), respectively. The excitation frequency is maintained at \( 1.1\,\mathrm{Hz} \), and the excitation amplitude is fixed at \( 0.01\,\mathrm{m} \).

\begin{figure}[htbp]
    \centering
    \begin{subfigure}{\textwidth}
        \centering
        \includegraphics[width=\textwidth]{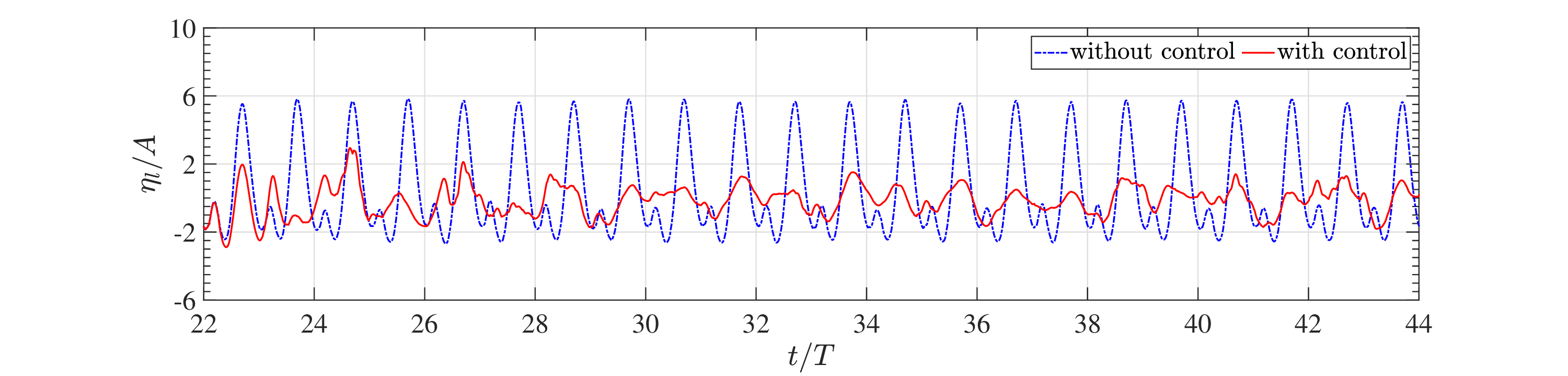}
        \caption{\(0.15 \, \mathrm{m}\)}
    \end{subfigure}
    
    \begin{subfigure}{\textwidth}
        \centering
        \includegraphics[width=\textwidth]{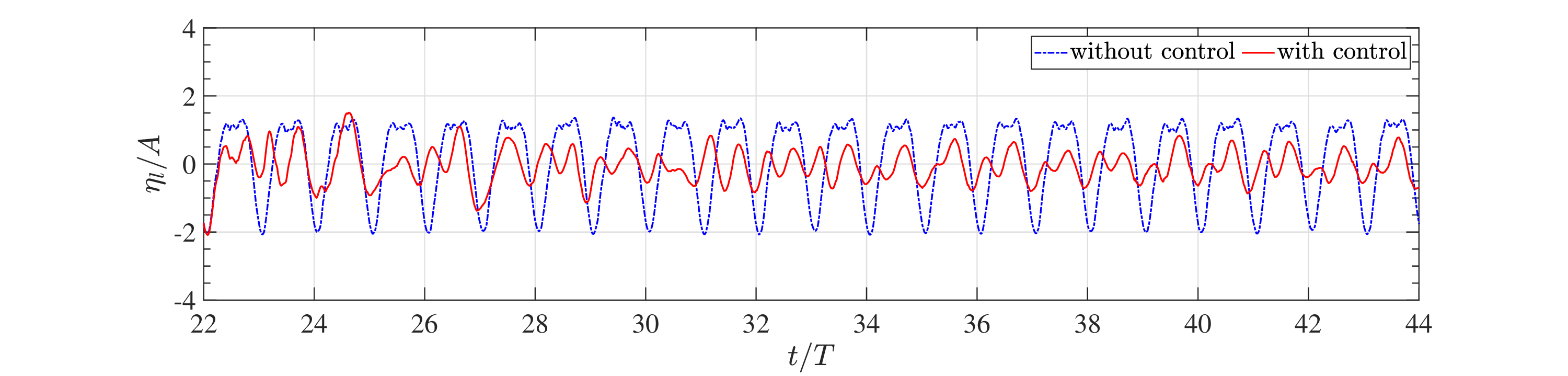}
        \caption{\(0.25 \, \mathrm{m}\)}
    \end{subfigure}

    \caption{Time history of the free surface height at the left wall under different water depths: (a) \(0.15 \, \mathrm{m}\), (b) \(0.25 \, \mathrm{m}\).}
    \label{fig:22}
\end{figure}

As shown in Fig.~\ref{fig:22}, the lateral motion of the baffle significantly suppresses sloshing across different water depths. Specifically, for a water depth of \(0.15\,\mathrm{m}\), the mean free surface elevation \(\eta_{\mathrm{mean}}\) is reduced from \(0.0218\,\mathrm{m}\) to \(0.0063\,\mathrm{m}\), corresponding to a 71.1\% reduction. In the case of \(0.25\,\mathrm{m}\) water depth, \(\eta_{\mathrm{mean}}\) decreases from \(0.011\,\mathrm{m}\) to \(0.0033\,\mathrm{m}\), yielding a 70\% reduction.

\begin{figure}
\centering
\includegraphics[width=\textwidth]{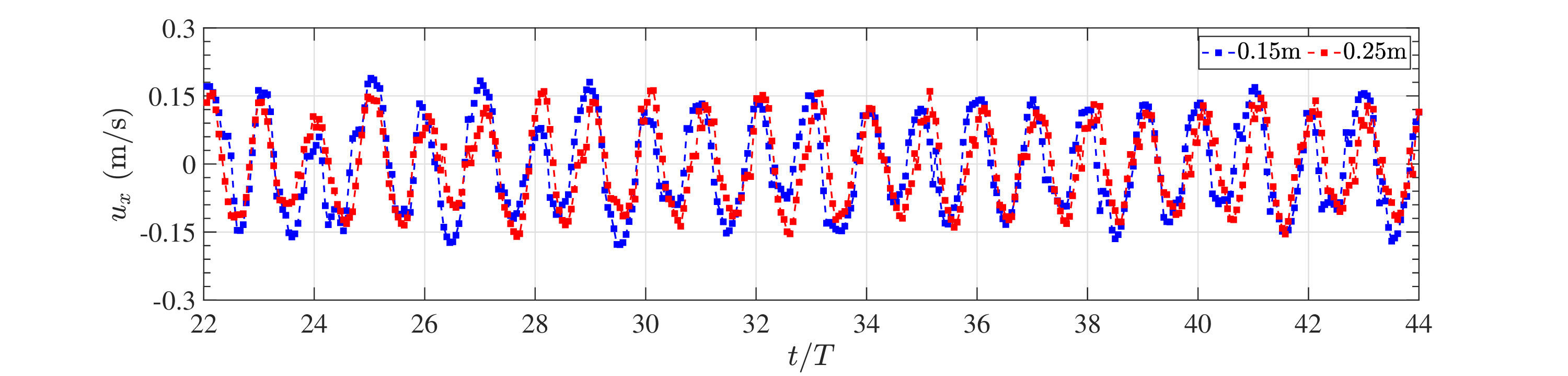}
\caption{Time history of the baffle’s velocity at different water depths.}
\label{fig:23}
\end{figure}

\begin{figure}[htbp]
    \centering
    \begin{subfigure}{\textwidth}
        \centering
        \includegraphics[width=\textwidth]{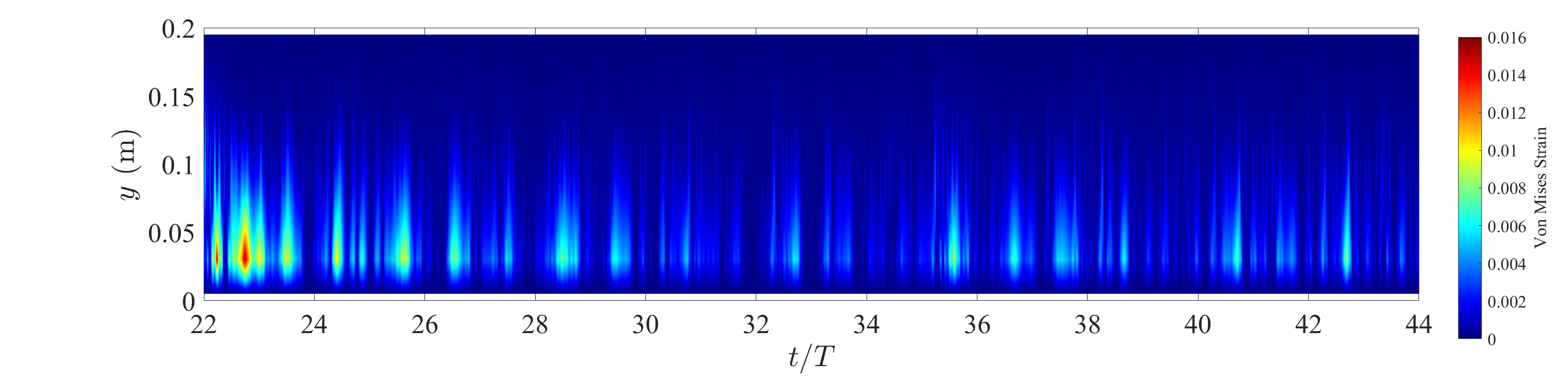}
        \caption{\(0.15 \, \mathrm{m}\)}
    \end{subfigure}
    
    \begin{subfigure}{\textwidth}
        \centering
        \includegraphics[width=\textwidth]{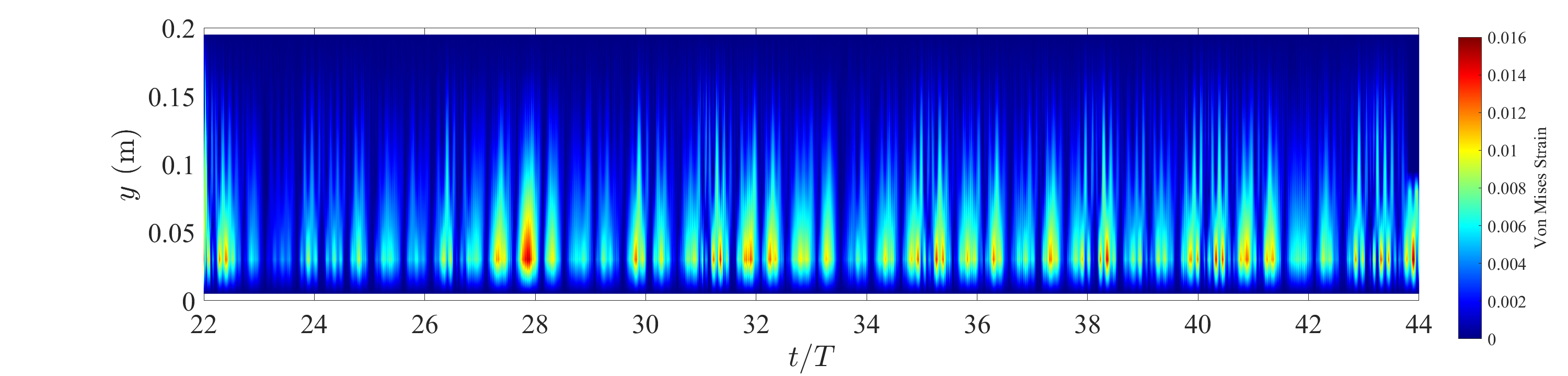}
        \caption{\(0.25 \, \mathrm{m}\)}
    \end{subfigure}

    \caption{Time history of the Von Mises strain distribution along the baffle height under different water depths: (a) \(0.15 \, \mathrm{m}\), (b) \(0.25 \, \mathrm{m}\).}
    \label{fig:24}
\end{figure}

Analysis of the baffle velocity profiles indicates that water depth has limited influence on the motion strategy. Even though the excitation frequency at \(0.15\,\mathrm{m}\) is closer to the system's natural frequency—leading to higher sloshing amplitudes (nearly double that of the \(0.25\,\mathrm{m}\) case)—the velocity amplitude remains approximately constant at \(A_b \approx 0.15\,\mathrm{m/s}\), as shown in Fig.~\ref{fig:23}. Although the motion at the base of the baffle is similar in both cases, the overall deformation of the baffle differs significantly. As illustrated in Fig.~\ref{fig:7} (d) and Fig.~\ref{fig:24}, increasing water depth results in greater hydrodynamic resistance during baffle movement, especially increasing strain in the lower half (approximately from \(0.025\,\mathrm{m}\) to \(0.05\,\mathrm{m}\) in height). Notably, the bending direction under active control is opposite to that observed in the uncontrolled case, and the strain magnitude is considerably smaller.

\begin{figure}[htbp]
    \centering
    \begin{subfigure}{\textwidth}
        \centering
        \includegraphics[width=\textwidth]{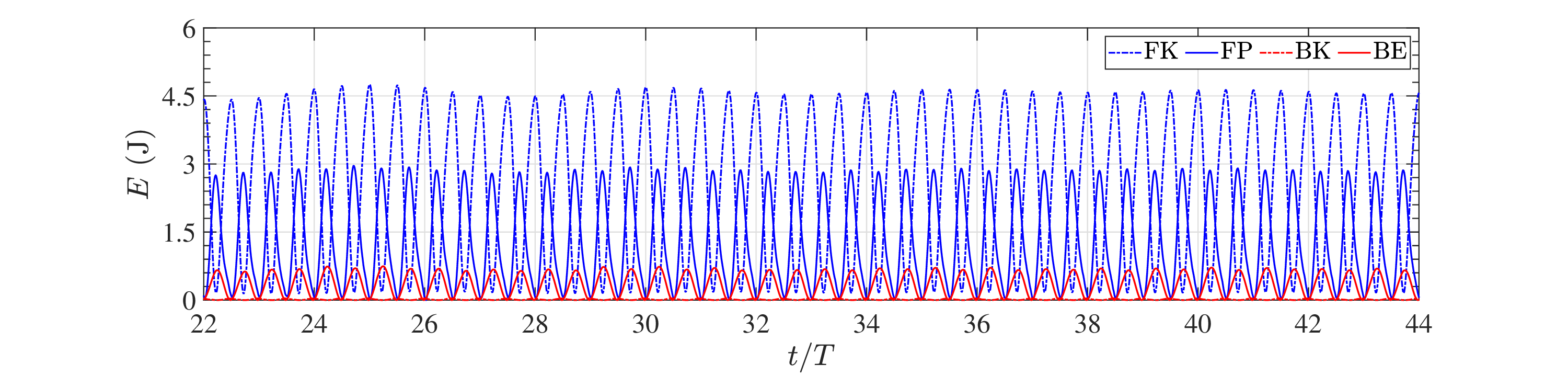}
        \caption{baffle motion only}
    \end{subfigure}
    
    \begin{subfigure}{\textwidth}
        \centering
        \includegraphics[width=\textwidth]{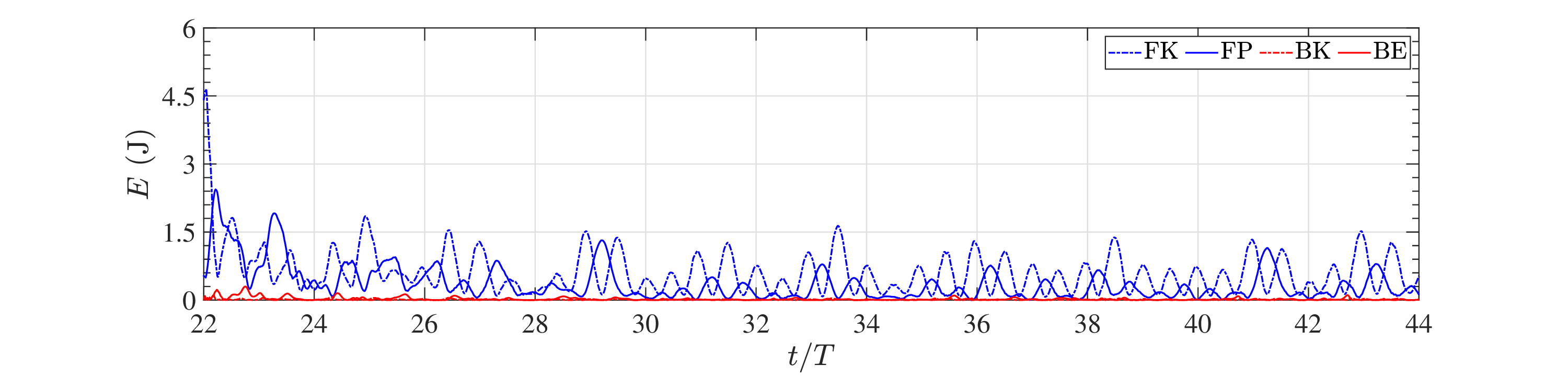}
        \caption{without control}
    \end{subfigure}

    \begin{subfigure}{\textwidth}
        \centering
        \includegraphics[width=\textwidth]{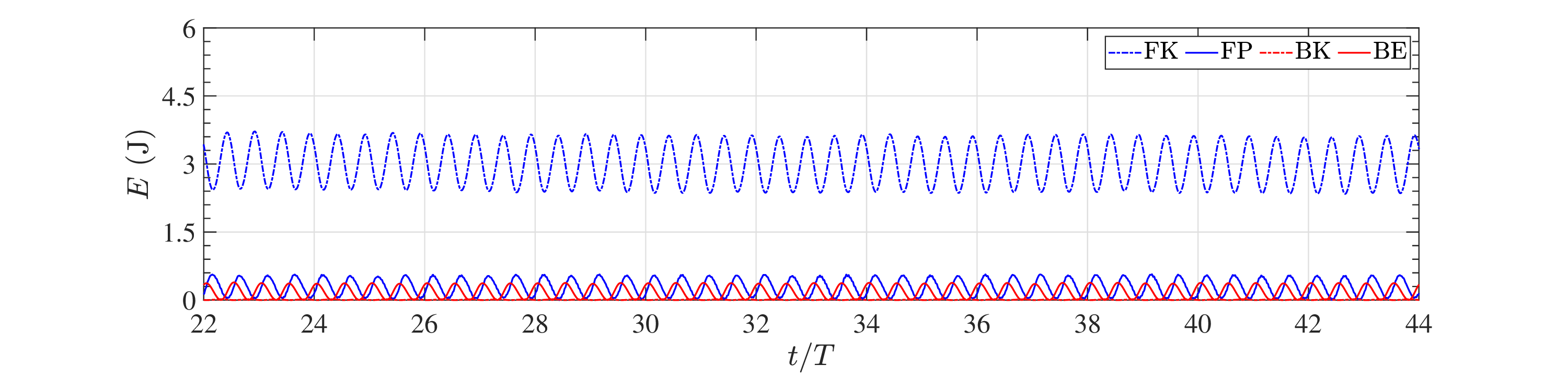}
        \caption{with control}
    \end{subfigure}

    \begin{subfigure}{\textwidth}
        \centering
        \includegraphics[width=\textwidth]{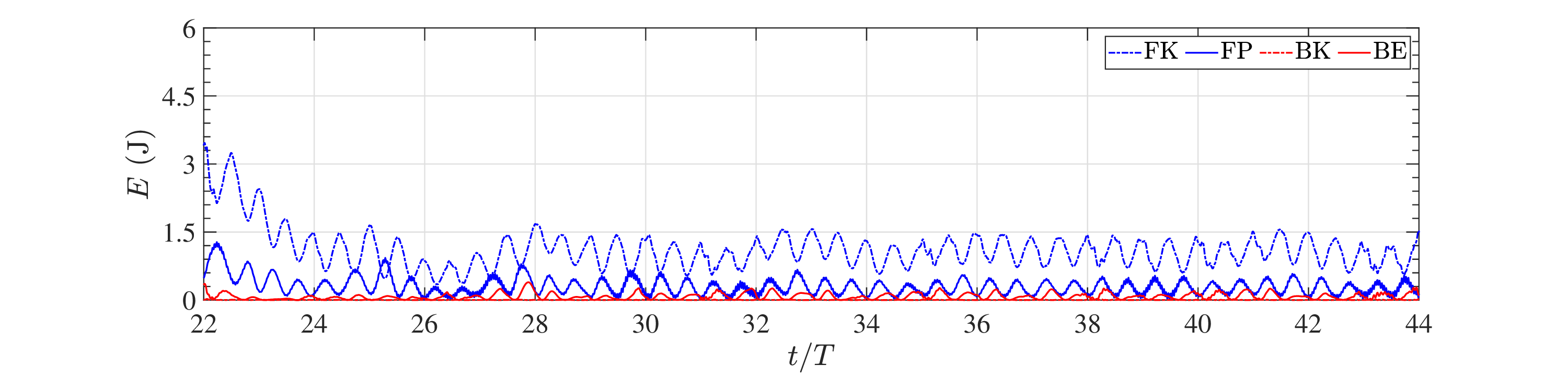}
        \caption{with control}
    \end{subfigure}

    \caption{Time histories of FK, FP, BK and BE under different water depths: (a) (b) \(0.15 \, \mathrm{m}\), (c) (d) \(0.25 \, \mathrm{m}\).}
    \label{fig:25}
\end{figure}

Further analysis of the elastic energy evolution in Fig.~\ref{fig:11} (c) and Fig.~\ref{fig:25} reveals that for water depths of \(0.15\,\mathrm{m}\) and \(0.2\,\mathrm{m}\), the elastic energy of the baffle nearly vanishes during motion. This indicates that the elastic baffle behaves similarly to a rigid baffle in these cases. In contrast, for \(0.25\,\mathrm{m}\) depth, the elastic energy does not diminish during control, highlighting the importance of deformation-induced shifts in the tank natural frequency. Moreover, while potential energy remains unchanged, the observed suppression is primarily attributed to a reduction in fluid kinetic energy.

\section{Conclusions}\label{Conclusions}
This study employed a DRL-based control framework, integrating the numerical platform SPHinXsys with the DRL platform Tianshou to effectively mitigate fluid sloshing in tanks with elastic baffles. Under baseline conditions ($h = 0.2\,\mathrm{m}$, $f_e = 1.1\,\mathrm{Hz}$), the trained DRL policy achieved an 81.63\% reduction in mean free surface amplitude. Additionally, a cosine-based expert policy derived from the DRL velocity profile demonstrated comparable effectiveness, yielding a 76.86\% reduction, confirming DRL-informed policies' practical applicability.

Comparisons among control strategies showed that moving rigid and elastic baffles significantly attenuate sloshing, with the elastic baffle consistently delivering superior performance. In contrast, the active strain method generated stronger local disturbances and proved less effective. Energy analysis confirmed that active baffle motion performed negative work on the fluid, substantially reducing the system's kinetic and potential energy, thus highlighting the robustness and efficiency of displacement-based control approaches, particularly elastic baffle motion.

The DRL-based control maintained high suppression performance across varying excitation frequencies. At lower frequencies ($f_e = 0.9\,\mathrm{Hz}$), rigid baffles were slightly more effective, achieving a 72.0\% reduction compared to 68.75\% with elastic baffles. Conversely, at higher frequencies ($f_e = 1.3\,\mathrm{Hz}$), elastic baffles outperformed rigid ones, achieving reductions of 71.43\% versus 64.85\%. Cosine-based expert policies derived from DRL further improved suppression effectiveness, confirming that structural flexibility significantly enhances sloshing control under diverse nonlinear FSI conditions.

Finally, simulations across different water depths ($h = 0.15\,\mathrm{m}$ and $0.25\,\mathrm{m}$) demonstrated that the elastic baffle's lateral motion effectively suppressed sloshing, achieving amplitude reductions of approximately 70\% in both cases. Although the control policy proved insensitive to changes in water depth, deformation behavior notably differed, with deeper water inducing greater hydrodynamic resistance and structural deformation. 

Future work will explore the effectiveness of DRL-based baffle control under more complex external excitation conditions, such as irregular wave patterns or rotational motion. This will provide further insights into whether DRL policies can effectively suppress sloshing in scenarios closer to practical applications. The proposed control policies also require experimental validation to confirm their applicability and robustness.

\appendix
\setcounter{figure}{0}
\section{Convergence analysis of the number of state sensors}\label{app1}
Here, the rigid baffle is employed to suppress sloshing at a water depth of 0.2 m by controlling its lateral motion. The model's physical parameters remain consistent with those of the water tank with an elastic baffle. The external excitation parameters are set to \( A = 0.01 \, \mathrm{m} \) and \( f_e = 1.1 \, \mathrm{Hz} \).

\begin{figure}
  \centerline{\includegraphics[scale=0.65]{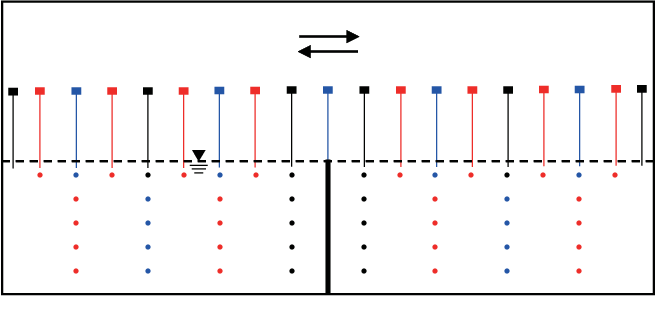}}
  \caption{Distribution of velocity probes (in \(x\)- and \(y\)-directions) and free-surface height probes. Black probes represent the distribution used for 32 observations; blue probes indicate additional probes added to the black distribution to reach 61 observations; red probes denote additional probes included on top of the black and blue distributions.}
\label{fig:ap1}
\end{figure}

To assess the independence of the number of observations, we conducted a sensitivity analysis, as illustrated in Fig.~\ref{fig:ap1}. For example, in the 61-observations configuration, the placement includes:
\begin{enumerate}
     \item 24 velocity probes, uniformly distributed at: 
     \[x = \{0.2, 0.4, 0.6, 0.8\} \text{ m}, \quad y = \{0.02, 0.06, 0.1, 0.14, 0.18\} \text{ m}.\] 
     \item 11 free surface height probes, including measurements at the left and right walls, with 0.1 m spacing between adjacent probes.
     \item Two additional probes monitoring the velocity and position of the baffle.
     \item  For the elastic baffle, four additional probes are attached to the baffle to monitor its deformation, while the baffle velocity is determined by tracking the deformation-free region at its base.
\end{enumerate}

\begin{figure}[htbp]
    \centering
    \begin{subfigure}{0.48\textwidth}
        \centering
        \includegraphics[width=\linewidth]{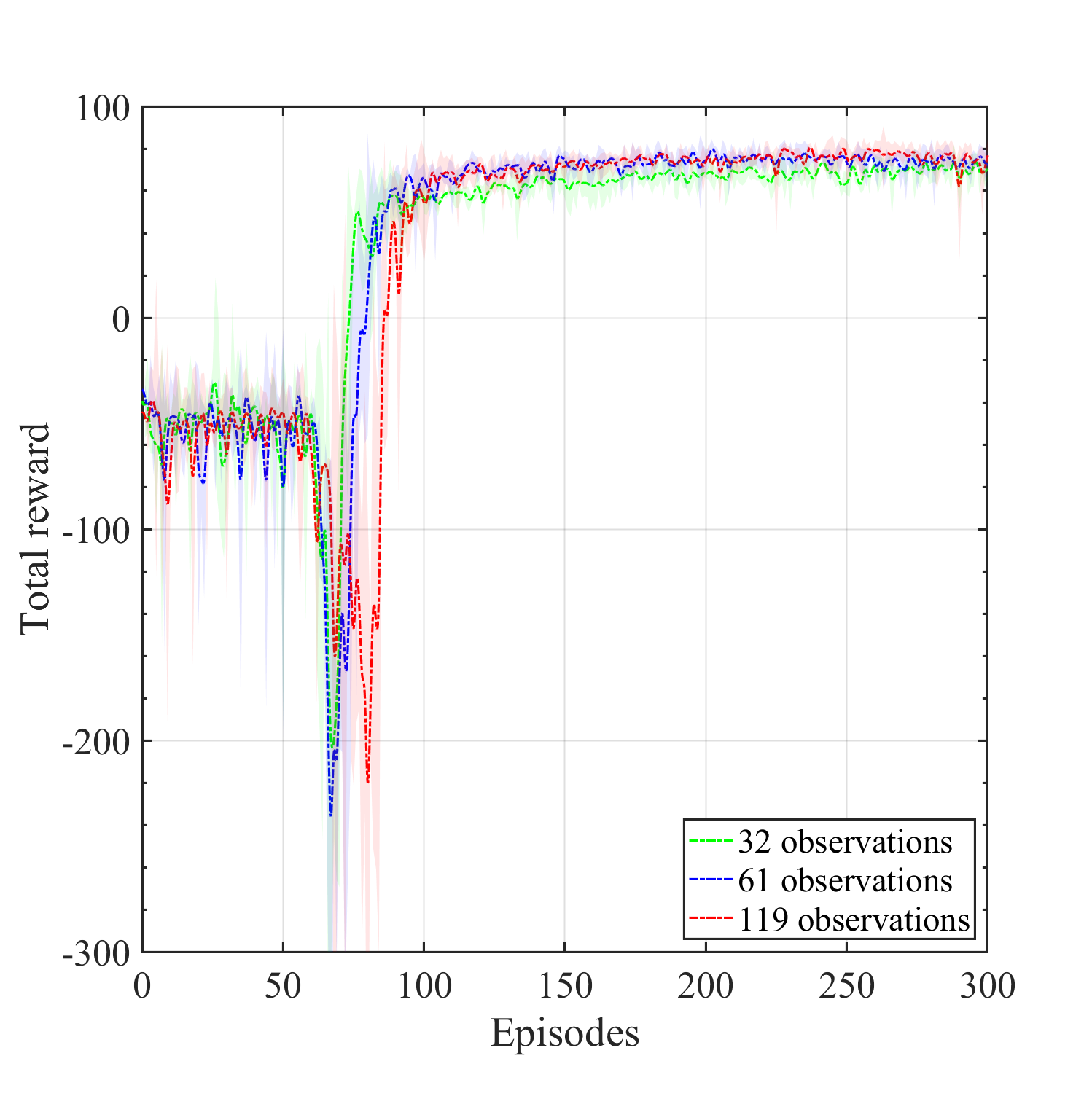}
    \end{subfigure}
    \hfill
    \begin{subfigure}{0.48\textwidth}
        \centering
        \includegraphics[width=\linewidth]{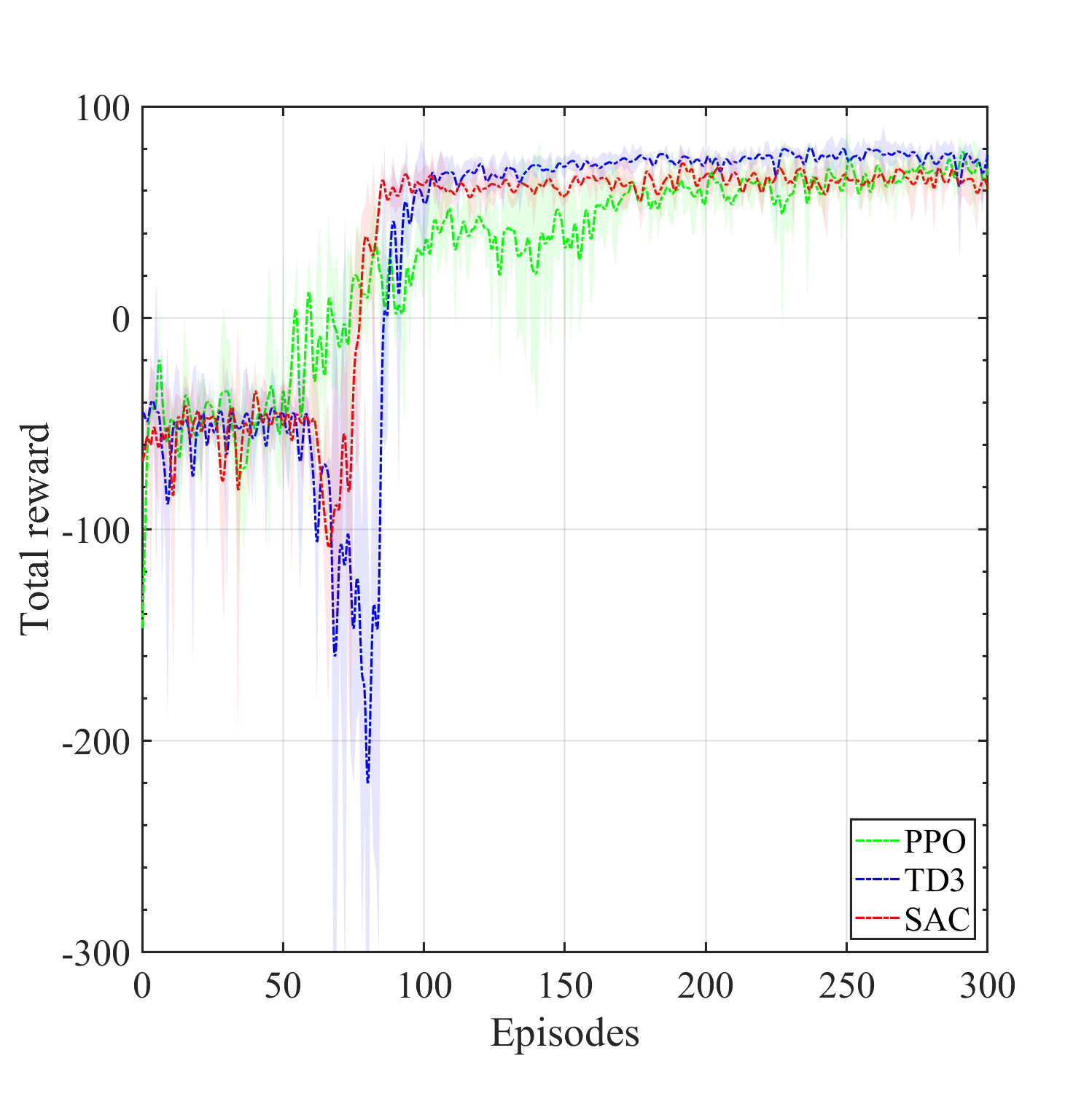}
    \end{subfigure}

    \caption{Effects of observations and DRL algorithms on sloshing suppression performance.}
    \label{fig:ap2}
\end{figure}

As shown in Fig.~\ref{fig:ap2} (a), increasing the number of observation points improves the effectiveness of the control policy. However, the results obtained using 61 and 119 observations are highly similar, and the policy trained with 119 observations exhibits a significantly slower convergence speed than 61 observations. Therefore, in this study, we adopt 61 observations for the rigid baffle and 65 for the elastic baffle as the state representation at a given time step. Furthermore, as shown in Fig.~\ref{fig:ap2} (b), the TD3 algorithm converges faster than the PPO algorithm but slower than the SAC algorithm while achieving the best overall performance. Although the TD3 algorithm exhibits significant fluctuations during the initial training phase following the pre-training data collection, this behavior is primarily attributed to its use of direct action output rather than a probability density function. Additionally, introducing random noise substantially enhances the algorithm’s exploration capability.

\section*{CRediT authorship contribution statement}
\textbf{M. Ye:} Validation, Methodology, Investigation, Formal analysis, Writing - original draft, Writing - review \& editing. \textbf{Y.R. Ren:} Methodology, Writing – review \& editing, Investigation. \textbf{S.L. Zhang:} Resources, Conceptualization. \textbf{H. Ma:} Methodology, Data curation. \textbf{X.Y. Hu:} Writing – review \& editing, Supervision, Conceptualization. \textbf{O.J. Haidn:} Supervision, Investigation.

\section*{Declaration of competing interest}
The authors declare that they have no known competing financial interests or personal relationships that could have appeared to influence the work reported in this paper.

\section*{Data availability}
The corresponding data of this work will be made available on reasonable request.

\section*{Acknowledgments}
M. Ye was supported by China Scholarship Council (No.202006120018) when he conducted this work.

\bibliographystyle{elsarticle-num-names} 
\bibliography{baffle}

\end{document}